\documentclass[12pt,preprint]{aastex}

\newcommand{\Msun}{\rm M$_\odot$}

\newcommand{\Rsun}{\rm R$_\odot$}
\newcommand{\Rearth}{\rm R$_\oplus$}

\shorttitle{Time Series UV Photometry of White Dwarfs} 
\shortauthors{Debes et al.}
\begin{document}
\title{A Search for short-period Rocky Planets around WDs with the Cosmic Origins Spectrograph (COS)\footnote{Based on observations made with the NASA/ESA Hubble Space Telescope, obtained from the data archive at the Space Telescope Science Institute. STScI is operated by the Association of Universities for Research in Astronomy, Inc. under NASA contract NAS 5-26555.}}
\author{Phoebe H. Sandhaus\altaffilmark{1,2}, John H. Debes\altaffilmark{2}, Justin Ely\altaffilmark{2}, Dean C. Hines\altaffilmark{2}, Matthew Bourque \altaffilmark{2}}

\altaffiltext{1}{The Ingenuity Project}
\altaffiltext{2}{Space Telescope Science Institute, 3700 San Martin Dr., Baltimore, MD 21218}

\begin{abstract}
The search for transiting habitable exoplanets has broadened to include several types of stars that are smaller than the Sun in an attempt to increase the observed transit depth and hence the atmospheric signal of the planet.  Of all spectral types, white dwarfs are the most favorable for this type of investigation.  The fraction of white dwarfs that possess close-in rocky planets is unknown, but several large angle stellar surveys have the photometric precision and cadence to discover at least one if they are common.  Ultraviolet observations of white dwarfs may allow for detection of molecular oxygen or ozone in the atmosphere of a terrestrial planet.  We use archival Hubble Space Telescope data from the Cosmic Origins Spectrograph to search for transiting rocky planets around UV-bright white dwarfs.  In the process, we discovered unusual variability in the pulsating white dwarf GD~133, which shows slow sinusoidal variations in the UV.  While we detect no planets around our small sample of targets, we do place stringent limits on the possibility of transiting planets, down to sub-lunar radii.  We also point out that non-transiting small planets in thermal equilibrium are detectable around hotter white dwarfs through infrared excesses, and identify two candidates.
\end{abstract}


\section{Introduction}
The successful discovery of planets with $\sim$\Rearth\ radii and total stellar insolation similar to the Earth via the {\em Kepler} mission \citep[e.g.][]{jenkins15} has motivated an intense effort to lay the groundwork for spectroscopic determinations of terrestrial exoplanet atmospheres via transit spectroscopy.  Currently giant planet and super-Earth atmospheres are accessible via space- and ground-based observatories, but searching the atmospheres of rocky planets for water in the so-called habitable zone of their stars will be challenging if not impossible for even the high sensitivity of JWST.  However, the robustness of planet formation, from a planetary system in orbit around the pulsar PSR 1257+12 \citep{wolszczan92} to planetary mass companions around brown dwarfs \citep{chauvin05}, implies that broader searches for terrestrial mass planets may yield a more favorable radius ratio for transit spectroscopy investigations.

One of the most favorable combinations of transit probability, radius ratio, and orbital period stem from white dwarfs (WDs) \citep{agol11}.  WDs mark the end of stellar evolution for stars with initial masses from 0.07 to 8~\Msun, with over 97\% of main sequence stars becoming WDs \citep{fontaine01}.  Their current space density is similar to that of G-type stars and, after solar-type and M-dwarfs, represent the third most common stellar class to search for planetary mass objects \citep{holberg08}.

Although white dwarfs no longer have a source of thermonuclear energy, they can support a potentially habitable zone (PHZ) for billions of years.  The newly formed white dwarfs, over a hundred thousand degrees Kelvin, push an instantaneous PHZ to much greater distances.  However, as the white dwarf quickly cools to a temperature similar to that of the sun, the PHZ moves dramatically inward and certain orbital radii are continuously in the PHZ for over 1~Gyr \citep{agol11}.  The small orbital radii of continuous PHZs for WDs increase the transit probability of terrestrial mass planets.  The transit probability of a terrestrial planet in the continuous PHZ is of the order of 1\%, and thus several thousand WDs will need to be observed to determine $\eta_\oplus$ for WDs \citep{agol11}.  Such surveys are currently possible via the Transiting Exoplanet Survey Satellite \citep{ricker15}, K2, and multiple ground-based large sky survey efforts like Evryscope \citep{law15} and LSST \citep{lund15}, which are sensitive to $\sim$30\% transit-depth signals on short cadences down to apparent optical magnitudes of $\sim$16-17 or greater.  The local volume of such WDs will yield transiting planets, transiting brown dwarfs that survive post-common envelope evolution \citep{casewell12}, variability due to pulsations, variability due to the magnetic and spot properties of WDs \citep{holberg11}, multiple eclipsing WD binary systems \citep[e.g.][]{hallakoun15}, and tidal disruptions of asteroids in real time \citep{bear13, vanderberg15, croll15}.

The secure detection of an atmosphere of a terrestrial planet orbiting a white dwarf will require careful multi-wavelength follow-up.  Optical and NIR photometric precisions are in principle sufficient to very accurately characterize apparent planet radius with wavelength, and the James Webb Space Telescope (JWST) will also be sensitive to certain atmospheric constituents, including the byproducts of certain industrial processes \citep{loeb13,lin14}.  Earth analogs will have a strong signature of O$_2$ and O$_3$ in their atmospheres, which is a potential indicator of conditions amenable to life.  The strongest signatures of these atmospheric products are in the FUV and NUV \citep{betremieux13}.  The Hubble Space Telescope (HST), equipped with the Space Telescope Imaging Spectrograph \citep{woodgate98} and the Cosmic Origins Spectrograph \citep{green12}, is the only currently operating observatory with NUV and FUV spectroscopic capabilities.  

In this paper we investigate the ability of COS, the most sensitive point source UV-spectrograph on HST, to be utilized as a high speed photometer.  Throughout the course of COS' lifetime, it has observed over 100 WDs, some with total exposure times exceeding that which is needed to search for PHZ WD planets.  In \S \ref{sec:wdlc} we detail our method of extracting high quality lightcurves from COS spectra and investigate the instrument's capability to follow-up and characterize any planets discovered.  In \S \ref{sec:search} we detail our initial search of WDs with archival COS observations for short-period variability and transiting planets, including the observation of UV pulsations in previously discovered ZZ Ceti stars.  We also discover, for the first time, low amplitude, long period variability in GD~133 that is potentially due to a spot on its surface.  We then determine our sensitivity to transiting rocky bodies for seven WDs that had total exposure times $>$7200~s.  In \S \ref{sec:limits2} we determine that WDs with T$_{\rm eff}>15000~K$ can show IR-excesses from progenitors of PHZ planets irrespective of whether they transit or not.  We therefore place limits to non-transiting close-in planets around our sample of COS WDs and also determine two candidate close-in planetary WD systems.  In \S \ref{sec:disc} we present the implications of our work.

\section{White Dwarf Light Curves with COS}
\label{sec:wdlc}

COS was designed as a high throughput spectrograph focused on extreme sensitivity.  Its far-ultraviolet (FUV) mode utilizes two independent crossed-delay line micro-channel plate photon counting devices, commonly referred to as Segment A and Segment B \citep{green12}.  For the near-ultraviolet (NUV) modes, COS has a single 1024$\times$1024 Multi-Anode Micro-channel Array.  UV photons that strike either the NUV or FUV detectors create clouds of electrons that enable the COS instrument to record the location and time of each photon.  This method of data-taking for COS is called the TIME-TAG mode.  We use this feature of COS to convert archival TIME-TAG spectral observations of COS into lightcurves.  In contrast, the standard calibration pipeline (CalCOS) takes the same raw data, corrects for instrumental effects, extracts one-dimensional spectra, and produces a final time-integrated, flux calibrated spectrum.

As part of the Archival Legacy Program 13902 ``The Lightcurve Legacy of COS and STIS'' (PI: Ely), we have developed calibration routines that take corrected TIME-TAG data and extract time resolved count rates for FUV and NUV data over selectable wavelength ranges.  In this Section, we describe the process by which lightcurves are extracted, expected countrates for a range of white dwarfs, and the overall stability and repeatability of COS photometry.

\subsection{Extraction of Time Series Photometry from COS spectra}
\label{sec:wdlc:extract}

Both the FUV and NUV detectors on COS observe in TIME-TAG mode by default - meaning that each photon's arrival time is recorded down to 32~ms precision.  The CalCOS pipeline applies certain corrections in a time-dependent fashion to these events, but eventually bins them to a two-dimensional image in order to calibrate and extract a one-dimensional spectrum.  The extraction code employed here instead filters, calibrates, and bins these events in time to create a time-series lightcurve.

Nominally, events are first filtered out if they fall outside the standard extraction boxes, hit bad-pixel-regions of the detector, or fall in airglow windows.  Additional filtering can be done according to user-selectable wavelength and detector position parameters to the extraction.  The flatfield and deadtime correction factors are then used as weights to a histogram routine to extract a gross count-rate lightcurve.  For the flux-calibrated light curve, the time-dependent sensitivity and flux-calibration reference files are re-factored from their standard form to provide an appropriate correction factor to each individual event.  These corrections are then used to re-extract the lightcurve in flux-space.  A background light curve is also extracted using the same procedure as for the source counts, but using pre-defined background regions above and below the spectral extraction window.  The background light curve is used to do background subtraction of the final product.  A final FITS table of counts, count rates, background, and flux versus time is produced for each input exposure.    

\subsection{Feasibility Calculations for White Dwarf Lightcurves}
\label{sec:wdlc:etc}

COS is the most sensitive point source UV spectrograph currently in space.  Here we determine how feasible it will be to conduct targeted follow-up of WDs with high quality time series photometry.  In general, given the large wavelength coverage of the COS gratings, high S/N photometry is more feasible for a given source than moderate resolution spectroscopy.  We first compared the expected count rates for selected COS modes for two COS WD standards, GD 71 and WD 0308-565.  WD 0308-565 was observed  with the G130M/1291 grating and central wavelength combination on 8 June 2015 as part of routine sensitivity monitoring for COS.  The median count rate for the 224s exposure was 6920 cts/s over the entire spectral range.  GD 71 was observed with the G160M/1577 grating and central wavelength on 13 April 2015 also for routine sensitivity monitoring and with Segment B of the instrument turned off.  The median count rate for the 102s exposure was 7700~cts/s.  We compare the observed count rates to those given by version 23.2 of the STScI exposure time calculator (ETC), which predicts COS performance specifically for the time period of March 2016.  This date is important, since COS' gratings suffer from time dependent sensitivity degradation at the rate of 4-10\%/yr \citep{osten11}.  While the details of the degradation can vary, the ETC is generally accurate to 5-10\% for COS observations that occur within 1 year of the target calculation date.  We determine the expected source count rate by retrieving the pixel-to-pixel ETC results and summing over the pixels illuminated by the source in the appropriate wavelength range.  For the G130M/1291 setup and WD 0308-565, we obtain a predicted count rate of 7271 cts/s, roughly 5\% higher than observed.  For GD~71, we obtain a predicted count rate of 7160 cts/s, about 7\% lower than observed.  These differences are within the expected uncertainties of the calculations given by the ETC and show that time series photometry observations with COS can be planned effectively with the existing ETC.

We extend this analysis for a broader range of WD effective temperature to investigate the limiting V magnitude for S/N=10 time series photometry assuming 30s sampling, with the above caveat that these values will change slightly with time given the details of the sensitivity degradation of COS.  However, since we show below in Section \ref{sec:wdlc:stable} that COS is reasonably well described by photon-limited counting statistics up to near the bright limits of the detectors, additional sensitivity can be gained if a transiting system is known by integrating over multiple transits.  

Model hydrogen atmosphere white dwarf SEDs were input into the COS ETC for FUV and NUV spectroscopy, as well as NUV imaging, and limiting magnitudes were calculated and are summarized in Table \ref{tab:etc}.  In general, terrestrial planet transits around WDs can be confirmed or further characterized in the UV out to magnitudes comparable to many existing visible transit searches, but the choice of COS instrument mode will be determined by the wavelength range of interest as well as the WD temperature.

In addition to WDs, COS will have sufficient sensitivity for any astrophysical object that has flux in the FUV or NUV.  The only limits to COS are the bright object limits, which for continuum dominated sources correspond to global count rates of 30,000 cts/s in the FUV and 45,000 cts/s in the NUV.  This presents a limit to the radius of a transiting source that can be detected with COS in a single epoch, assuming a 3-$\sigma$ detection:

\begin{equation}
R_{\rm limit}= 0.043 R_\star \left(\frac{45000}{{\rm counts s^{-1}}}\right)^\frac{1}{4}\left(\frac{60 {\rm s}}{\Delta t}\right)^\frac{1}{4}
\end{equation}

A-type stars as well as subdwarfs will also be good sources to obtain COS time series photometry, and will still potentially be sensitive to planetary radius companions.  

\subsection{Stability of COS Photometry}
\label{sec:wdlc:stable}

Before performing any analyses on WD data collected with COS, we looked into the potential systematic uncertainties of COS time-series photometry.  This investigation into the performance of the FUV detectors allows us to understand the uncertainty associated with each observation and the potential for systematic behavior.  The sample we used for this particular study consisted of roughly ninety different targets observed with the COS FUV detector with short exposure times, using lightcurves extracted on 20~s sampling.  A full list of our targets, the modes used, and their exposure times are tabulated in Appendix \ref{app:A}.  The median exposure time was ~9 min; however, exposure times ranged from as low as ~3 min to ~55 min.  Once the light curves had been extracted as described in \S \ref{sec:wdlc:extract}, we took the fractional RMS of the count rate observed over a single exposure, and compared our results to the predicted photon-limited fractional RMS (shown in Figure \ref{fig:cos}) \citep{gehrels86}.

The vast majority of observations in our data set had short exposure times and high count rates; a combination which produced fractional RMS values in line with that expected for the photon limit down to levels of 0.2\%.  Since all of our observations had high count rates (above 4,000 counts with 20~s sampling), the dark rate of the FUV detector did not significantly impact the precision observed.  However, the dark rate of the COS FUV detector is roughly 4.0$\times$10$^{-6}$ cnts s$^{-1}$ pixel$^{-1}$, and observations with low countrates will also have the dark rate as a significant source of noise in addition to that from just the target.  Over short time periods and for countrates approaching the bright-object limits, photometry is photon noise limited by the source alone.

Longer term photometric precision is more dependent on the calibration of the FUV grating sensitivities and their time dependent behavior.  It has been demonstrated that both the COS FUV and NUV grating sensitivities have been slowly declining since COS' installation during Servicing Mission 4 \citep{osten11}.  Therefore, the precision of long term photometry is impacted by the overall precision of the grating time dependent sensitivity calibration.  The sensitivity declines vary in slope over wavelength and the magnitude of the slopes is also time variable.  They are currently characterized with an accuracy of $\sim$2\% for any given observation.  COS observations within an orbit are not appreciably impacted by such declines.  For longer term variations, one would need to rely both on the flux calibration and time dependent sensitivity trends available as part of the standard COS pipeline-reduced products available through MAST.  The lightcurve extraction routine provides flux-calibrated wavelength integrated fluxes, but the calibrations are also based off of the standard COS flux-calibration and time-dependent sensitivity reference files and will suffer the same biases. 

To quantify how accurate the flux calibration is as a function of time, we have investigated the overall variation of the flux calibrated data for WD~0308-565 as a function of central wavelength for the G130M (1291, 1300, 1309, 1318, and 1327) and G160M (1577, 1589, 1600, 1611, 1623) gratings.  {The average fractional RMS in flux for these gratings is 2.4\% and 2.5\% over a total of 4.3 and 2.6 years respectively since the first observation of WD~0308-565 was obtained in 2011. 

We investigated intermediate timescales, such as over multiple orbits or exposures of the telescope, where more subtle systematics are observed.  Part of a typical COS observation is to obtain four separate exposures over slightly different wavelength ranges, or Fixed pattern positions (FP-POS).  An FP-POS setting is executed by moving the optical grating mechanism by one quarter the distance it normally moves for a change in central wavelength, which has the effect of smoothing over fixed pattern features on the COS detectors.  The combination of slightly different wavelength ranges per FP-POS setting and a target's spectral energy distribution creates systematic offsets between exposures.

From orbit to orbit, HST's pointing is stable, but can drift or stochastically change by up to a few milli-arcseconds--we define this as spacecraft jitter.  HST's focus also secularly changes by small amounts due to slightly differing thermal conditions within the telescope, which can cause the cross-dispersion centroid and shape of a COS spectrum on the detector to change subtly.  Finally, from orbit to orbit, there can be slight offsets in where the target falls within the COS aperture.  While our extraction algorithm uses a wide extraction box that should encompass the source flux, COS does have non-negligible scattered light in the wings of its cross-dispersion profile, such that drifts on the detector may be detected.  COSÕ aperture suffers from vignetting if the target is more than a few tens of pixels (corresponding to $\sim$0\farcs4) from the center of the aperture, which could affect the total observed count-rates at a low level for smaller shifts of 1-2 pixels.

We investigated the exposure-to-exposure and orbit-to-orbit systematics of COS lightcurves.  For this investigation, we looked at several G130M spectra obtained as part of the flux calibration activities during the move of the COS spectral location from the second to the third lifetime position as part of program 13932.  The target for this program was WD 0308-565 and the exposures for these modes spanned several orbits, so the program provides a closer look at slowly varying trends we would not observe with the shorter exposures investigated in Section \ref{sec:uvvar}.

We first investigated the centroid of the cross dispersion profile on the detector as a function of time as well as the jitter to see if this was correlated with any changes in the reported count rates for sections of the detector that overlap in all four FP-POS.  This is important at the 1-3\% level depending on the spectrum of the source--a sharply defined SED at the edges of the detector will create significant total count rate changes as a function of FP-POS setting.  For small changes in the centroid position, we saw no clear correlation with the count rate of a source, nor were there clear correlations with the spacecraft jitter for this target.  In Section \ref{sec:uvvar}, however we do observe correlations between HST focus changes and COS spectrum centroid location and systematic changes in the observed count-rates for another target.  Therefore, on short timescales the photometry of COS is photon-limited, but on longer timescales one must be more careful about subtle trends which may or may not be present.

\section{A Search for Potentially Habitable Rocky Planets in Orbit around White Dwarfs}
\label{sec:search} 

A subset of all the WDs that have been observed with COS have significant total exposure times such that they are sensitive to transits of close-in planets with orbits $<$ 30~hr.  In this Section we define a sample of COS-observed WDs that can be searched for planets.  We also discuss the detection of significantly variable WDs in our larger sample of WDs investigated as part of \S \ref{sec:wdlc:stable}.

\subsection{White Dwarfs with long total exposure times}
We chose seven non-variable WDs with publicly available data and with total exposure times that exceeded 7200~s.  Table \ref{tab:targets} has the fundamental properties of our sample stars, and in Appendix \ref{app:B} we tabulate the individual exposures used in our analysis.
 All of the exposures except for those of SDSSJ122859.93+104032.9 (henceforth: SDSSJ1228+1040) were primarily used in calibration programs for HST.  Of the seven, four were primarily used for either monitoring or flux calibration for the COS FUV and NUV detectors.  GD153 and G191-B2b were primarily used for COS NUV monitoring and calibration.  

 All of the values for effective temperature, log g, and mass in Table \ref{tab:targets} were taken from \citet{gianninas11} with the exception of those for SDSSJ1228+1040 \citep{gaensicke12} and WD 0308-565 \citep{voss07}.  The instantaneous PHZ was calculated solely by incident stellar radiation, so that our assumptions were not dependent on atmospheric composition, but rather on the insolation of Mars and Venus.  In all cases, the orbital periods of planets in the instantaneous PHZs for our targets are much greater than the total exposure time, since the T$_{\rm eff}$ of our targets is so high.  Over the next few hundred million years, these WDs will cool such that their PHZ will encompass the periods where we searched for transiting planets, from 4-30~hr.  This range of periods corresponds to the continuous habitable zone for white dwarfs as defined in \citet{agol11}.

Most of the stars have pure hydrogen or helium atmospheric compositions and none of them possesses stellar companions.  G191-B2b has photospheric metal lines due to Mg II, Si II, and Fe II as first discovered by \citet{bruhweiler82}.  
SDSSJ1228+1040 posesses a metal-rich gaseous and dusty disk with both Mg II and Ca II emission lines in the optical \citep{gaensicke06,brinkworth09}.  The circumstellar disk has an outer radius of 1.2~\Rsun, and a high orbital inclination of  $>$70$^{\circ}$\citep{gaensicke06}.  Additional UV spectra have since shown additional metal lines due to C, O, Al, Si, and Fe within the WD's atmosphere\citep{gaensicke12}.  

\subsection{Light Curve Analysis}
\label{sec:wdlc:analysis}

          Before beginning the primary analysis of the COS WD data, we ran observations through the algorithm described in detail in \S \ref{sec:wdlc:extract} with 15~s sampling, and verified that there was no gross variability within each dataset.  In this process we detected one or two instances of data dropouts due to COS timing errors within observations that we subsequently excluded from our exposures.  Additionally, we removed all observations with exposure times of $<$60~s, since this is comparable to the timescale of the transits we are searching for.  Finally, for objects that had a large number of observations, we filtered out files that had larger standard deviations per normalized count-rate ($>$1\%) due to less sensitive modes.  Since SDSSJ1228+1040 and GD~153 had shorter total exposure times than the other targets, we did not filter low-sensitivity observations.

            Since our observations for each target could span multiple FP-POS settings, central wavelengths, detectors, and gratings, it was necessary to create flat lightcurves by normalizing the counts in each individual exposure by the median count-rate of each exposure.  After the normalization, we calculated the standard deviation of count-rates from the median count-rate of the exposure and searched for times where the observed count-rate varied by more than 5-$\sigma$ from the median.  We set our criterion to 5-$\sigma$ to ensure a low probability of a false positive in our large datasets, which exceeded tens of thousands of samples.  Given the heterogeneous nature of our datasets for each target, the criterion needed to be investigated for each individual exposure, rather than calculating a standard deviation across all normalized exposures.  

 We then folded the data over 4-30 hour periods and rebinned the observed count-rates in phase such that the smallest phase bin corresponded to 15~s.  Then, we determined the new uncertainty associated with each rebinned normalized count-rate ($\sigma_{\rm final}$) by taking the square root of the sum in quadrature of the original uncertainties of each point ($\sigma_{\rm i}$) divided by the number of points within the bin ($n$) squared.  This rebinning increased our sensitivity to periodic transits with stable ephemerides.  We again searched for any 5~$\sigma$ events in the rebinned, folded data.

            After completing this search,  we injected artificial planetary transits of varying depths and semi-major axes in order to quantify our sensitivity to transiting objects.  The inserted transits were modeled by the equation for a dark body transiting a uniform source \citep{mandel02}.  In order to imitate the fact that the transit could occur at any point in a given period, we randomly assigned the midpoint of the transit to a number between 0.0 and 1.0 which represented the phase at which the transit is at its greatest depth.  We repeated this randomization process one thousand times for each combination of period length between 4 and 30 hours (going in increments of one hour) and radius of the planet between 0.25 Earth radii and 2.0 Earth radii (going in increments of 0.25 Earth radii and accounting for when the planet was larger than the WD).  It was then recorded how many times out of the one thousand trials that the algorithm was able to detect the artificial transit.

             We created contour plots of our transit recovery percentage as a function of planet radius and period for each WD  (see Figures \ref{fig:0308} to \ref{fig:1228}.  Using the uncertainties of our rebinned and folded periods, we also calculated the optimistic and conservative detection limits for regularly periodic transit signals.  This was done by first looking at the uncertainties associated with each bin, and then picking out the maximum and minimum standard deviations at each period.  The maximum standard deviation would then serve as the conservative limit--the minimum as the optimistic limit. The 5$\sigma$ detection radii limits were expressed in kilometers and are also shown in Figures \ref{fig:0308} to \ref{fig:1228}, making the two assumptions that the object transiting the white dwarf is spherical in shape and revolving at an orbital inclination of 90$^\circ$.  The percentage recovery for these more sensitive limits are similar to those apparent in the contour plots of these figures.

\subsection{UV Variable White Dwarfs}
\label{sec:uvvar}
   In performing the analysis on the COS FUV detector to understand its sensitivity in \S \ref{sec:wdlc:stable}, two WDs were discovered to vary significantly in the ultraviolet on short timescales.  The two WDs, G29-38 and GD 133, both pulsate in the optical \citep{shulov74,mcgraw75,silvotti06}.  In addition to this, they both host circumstellar dust disks \citep{graham90,kilic06}.  Despite having total exposure times of $>$7200~s, we did not include them in our transit survey since they are shown to be significantly variable.

\subsubsection{G29-38}
G29-38's pulsations are relatively strong in the optical, with timescales that range around 600s and amplitudes that vary around 10\% \citep{shulov74,mcgraw75,kleinman98}.  Additionally, G29-38 has an infrared excess due to dust \citep{zuckerman87,graham90}, and shows significant photospheric pollution from the accretion of material from its dust disk \citep{koester97,xu14}.

            Archival TIME-TAG observations of G29-38 in the G130M/1300 grating exist from program GO 12290 (PI: Jura).  We observed large FUV variability in all four exposures we had available for the object, which corresponded to roughly 10,000s of total exposure time.  Figure \ref{fig:G29-38} illustrates one of these four observations, and clearly shows the distinct structure of each pulsation.  Some of the variations were double-peaked, while others had a single peak usually of larger amplitude.  These variations ranged in amplitude from $\sim$1.5-7.0 times the normalized median value of the continuum.  The pulsation periods also ranged in length from 3 to 11 min, within the range of periods observed in the visible.  A more detailed analysis of the UV pulsations is beyond the scope of this paper, but UV pulsation amplitudes can be used to better understand the exact mode of the pulsations \citep{kepler00}.

\subsubsection{GD 133}

GD 133 was first found to have calcium absorption features associated with its photosphere, and thus was classified as a DAZ \citep{koester05}.  It was also found to have a dusty disk \citep{kilic06}.   A short time later, \citet{silvotti06} discovered that GD 133 oscillates on very short time scales ($\sim$2 minutes) with relatively small amplitudes in the optical.  Using ten-second sampling, they detected peaks at 116 and 120 seconds, with amplitudes of 1.5 and 4.6 mma respectively.  After doing some additional analysis, \citet{silvotti06} found a possible third pulsation mode at 147 seconds with an amplitude of ~1.1 mma.  

The COS observations for GD~133 were taken as part of GO 12290 (PI: Jura), with results on the spectroscopy published in \citet{xu14} and \citet{xu2}.  The initial 15-s sampled lightcurves showed that the exposure-level fractional RMS on the lightcurves, $\sim$2\%, exceeded the expected noise by a factor of two.  We re-extracted a composite lightcurve of the full set of COS exposures using 1~s sampling of the lightcurve and with a wavelength range common to each FP-POS taken.  From this we calculated a Lomb-Scargle normalized periodogram via the IDL routine LN\_TEST, which is shown in Figure \ref{fig:lomb}.  A detailed analysis of the pulsations in GD~133 is beyond the scope of this paper, but we do report some main features of the UV pulsations.  Firstly, two strong peaks are observed in the periodogram, at periods of 146.6s (6.82~mHz) and 115.9s (8.62 mHz).  A less strong peak is also seen at 120.2s (8.32~mHz), but is comparable in strength to other peaks seen close to the 116s peak which are most likely due to the observing windows, which are roughly 50 minutes of every HST orbit.  All three of these periodicities were first reported by \citet{silvotti06}, where the amplitudes of these modes ranged from $\sim$5~mma for the 120~s mode in the $g$ band to 1.1~mma for the 146.6~s mode.  We folded the GD~133 lightcurve on the 146.6~s and 115.9~s periods and fit sinusoids, finding nearly equal amplitudes to the pulsations of 25~mma amd 24~mma respectively, factors of 23 and 5 higher than in the visible. 

In addition to the short-period pulsations, when we rebinned the lightcurve to 8 minute sampling to smooth over the pulsations, we detected significant variability at the level of $\sim$3\% peak-to-peak spanning the entire observation, implying a possible periodic variation of $\sim$5-6~hr duration.  The level of the variability is comparable to possible instrumental effects, so we investigated whether the observed flux was correlated with drifts in the position of the spectrum on the detector and with the modeled focus changes of HST over orbital timescales\footnote{http://www.stsci.edu/institute/org/telescopes/Reports/ISR-TEL-2011-01.pdf}.  It has been shown for other HST instruments that orbital variations in focus can affect what fraction of light is put into the wings of the instrumental PSF.  For COS, this could slightly change the amount of light in the PSF wings  which is vignetted within the point source aperture, but can also be caused by the truncation of light on the detector by our selection of an extraction box size for the photometry (See \S \ref{sec:wdlc:stable} for more details).

To investigate this behavior we re-extracted 120~s sampled lightcurves of the COS data from a narrow range of wavelengths (1340-1410\AA) to mitigate any countrate changes due to FP-POS effects.  We obtained a contemporaneous model\footnote{http://www.stsci.edu/hst/observatory/focus/FocusModel} of the HST focus behavior from STScI and constructed correlations between this model and the observed count rates in 120~s intervals.  From a linear fit to the correlation, we found a significant slope to the correlation and corrected the observed countrate.  For GD~133, we found that countrates varied by 0.5\% per \micron\ of focus change.  Over the course of the observations, the HST focus systematically started at +2\micron\ relative to the median focus and by the end of the orbit was -2\micron\ relative to the median.

After that step we correlated the corrected flux counts with the observed drift of the spectrum on the detector.  We measured the cross-dispersion drift by collapsing the spectrum on the detector in the calibrated CORRTAG files and by calculating an average spectral centroid with 120~s sampling.  For this calculation we used the XFULL and YFULL coordinates, since they should represent any uncorrected drifts by the internal calibration lamps of COS.  The XFULL and YFULL coordinates refer to the pixel locations in COS flatfield images and are corrected for geometric distortion and the wavelength dispersion.  Since the lightcurves for each target are extracted from a statically positioned box placed on top of the expected spectrum location, there is no accounting for slight mis-centerings of the target or of systematic drifts of the spectrum on the detector, which could affect the total amount of counts recorded for a given time.  We again find a significant correlation of count rate with drift offset and corrected the count rates for this.  The flux changes by 5\% per pixel of drift offset, and the spectra move on the detector between -0.2 to 0.1 pixels over the observations.

Figure \ref{fig:long} shows the resulting lightcurve sampled at 8-minute intervals and fit with a sinusoidal function with a period of 5.2~hr, an amplitude of 1.9\%, and T$_{\rm o}$ = 55709.279 MJD.  Sinusoidal variations in the UV and optical have been detected for a growing number of white dwarfs with periods similar to the timescale of these variations, many with observed photospheric metals \citep{holberg11,maoz,dupuis}.  In particular, the variations seen in the UV can be as much as 25\% \cite{dupuis}.  These features may be due to localized UV opacity on the WD surface \citep{dupuis,maoz}.  In the case of GD~133, which also possesses a dusty disk, the possibility exists that this variation is linked to the dust, possibly through an accretion hot spot.  We can estimate the accretion luminosity based on GD~133's inferred accretion rate of 2$\times$10$^7$ g~s$^{-1}$ and we get a luminosity that is too small by orders of magnitude compared to the amplitude of the variations.  Since the period is $>$4~hr, we can rule out a structure associated with the actual dust disk which is located at low orbital radii where the orbital periods are less than 4 hours.   We can rule out reflection or irradiation effects from planet-size objects as the amplitude is too large.  If this is due to UV opacity, there may be similar variations in the optical as well, but with potentially lower amplitude \citep{maoz}.

\subsection{Limits to Habitable Transiting Planets}
\label{sec:limits}
Due to their unique nature, WDs are prime objects to potentially observe a planetary transit.  Their small radii open possibilities for the detection of transiting objects much smaller than can be observed around main sequence stars.  These deep transits would also take place on very short timescales of ~1-2 min.  An example of a transit of a 0.5 Earth radius planet in a 6-hour period orbit around a WD is shown in Figure \ref{fig:tran}.  The insertion of the artificial transit was identical to the method described in detail in \S \ref{sec:wdlc:analysis}.  

Although all of the WDs in this study are extremely hot ($>$20,000 K), typically WDs quickly cool to well below $\sim$10,000 K in the first Gyr of their life, thus spending the vast majority of their time at much cooler temperatures.  In \citet{agol11}, the concept of the white dwarf habitable zone (WDHZ) is taken one step further to account for this lengthy temperature stability by creating a definition for the continuous habitable zone (CHZ).  The CHZ is defined as the portion of the habitable zone that would stay in the WDHZ for at least 3 Gyr.  The CHZ for the average 0.6 \Msun WD extends from $\sim$0.005-0.02 AU which corresponds to an orbital period range of 4-30 hrs \citep{agol11}. For sake of consistency in the analysis, this range of orbital periods was used to calculate the sensitivity to transiting planets around each WD, even though the WDs are currently too hot for the PHZs to correspond to this period range.  Eventually, these orbital periods will become the CHZ for each WD.

Based on our assumptions of spherical companions in orbits viewed edge on, we found that the greatest limiting factor to detecting transiting planets of a given radius was the total exposure time.  There is a very clear correlation between the total exposure time and the percent of the time that the algorithm detected the artificial transits.  This is primarily because there is a sharp transition between detection and non-detection of a source and because we are using time sampling that resolves the short transits of small planets.

Considering most of the WDs had very stable photometric observations, the optimistic detection limit in terms of radius was shown to be in many cases to objects smaller than Pluto.  The best case scenario for detection of a planetary transit (in terms of radius) was for the target WD 0947+857 (See Figure \ref{fig:0947}).  The optimistic limit for this target was calculated to be 526~km, while the conservative limit was 1232~km, making it possible for us to detect an object slightly larger than Ceres for this WD in the optimistic case, and slightly larger than Pluto in the conservative case.  In this particular case, the conservative limit is $\sim$0.6 times smaller than the smallest planet ever to be discovered via transit photometry, Kepler-37b \citep{barclay13}.  In all cases, the conservative limit was smaller than an Earth radius.

\section{Limits to Non-Transiting Planets}
\label{sec:limits2}
While WD transits are the most sensitive measure of rocky planets that have close-in orbits, we note that for hot WDs (T$_{eff} >$ 15000~K) slightly larger planets with small orbital semi-major axes in radiative equilibrium with their host white dwarf may also be observable and can mimic dusty circumstellar disks\citep{xu15}.  To investigate what kinds of planets can be detectable around hotter WDs via NIR excesses, we calculate the emission of a blackbody in radiative equilibrium with its host WD for a range of WD T$_{eff}$ with {\rm $\log~g=8.0$}, and assumed orbital radii that range between 1~\Rsun and 10~\Rsun. We use the predicted photometry of the WD from the cooling models of \citet{bergeron95} and assume photometry in 2MASS $J,H,Ks$ and WISE $W1, W2$.  We assume an albedo of 0.1 for the putative planet and that the planet is detectable once it shows an excess of $>$10\% in any of the above bandpasses (typically W2), which would correspond to a detection of 3$\sigma$ for a source with 3\% uncertainties in its photometry.  The range of temperatures probed roughly matches the maximum dayside temperatures of companions measured for highly irradiated brown dwarfs in orbit around WDs \citep{casewell15}, as well as hot Jupiters \citep{sing13,knutson12} and Super-Earths \citep{demory14} in orbit around main sequence stars.  Figure \ref{fig:superearth} shows the resulting limiting radii as a function of orbital separation and T$_{eff}$.  

We find that planets with radii of $>$2~\Rearth are detectable around hotter WDs.  A high precision IR search for WD planets (where photometric accuracy exceeds 1\%) might detect excesses from smaller planets as it orbits the primary.  Such close-in planets will most likely be tidally locked and show variability in the NIR due to the changing planetary phase \citep[e.g.,][but with larger amplitude]{lin14b}.  Since companions of $\sim$\Rearth have been proposed to explain long period variations in two pulsating subdwarf stars \citep{charpinet11,silvotti14}, a survey of a few hundred hot white dwarfs may turn up several candidates that could have their masses measured through radial velocity observations\citep{casewell12,charpinet11}.  

The detection of close-in planets will have implications for the process of common envelope evolution and the long term dynamical stability of tightly packed multi-planet systems.  From field RV measurements of planetary companions around solar-type stars \citep{cumming08}, and predictions of extreme orbital migration due to tidal evolution with a giant star for all Jupiters interior to $\sim$6~AU \citep{nordhaus13}, more than 10\% of WDs may have had a giant planet engulfed by the star during its red giant branch or asymptotic branch evolution.  Remnants of that process may survive to orbit the WD.  A large fraction of stars appear to house tightly packed multi-planet systems with super-Earths--some fraction of these may deliver, through dynamical instabilities, close-in companions that are heated during the early phases of WD cooling \citep{veras15}.  

Discovering irradiated WD planets will also be essential for understanding the potential habitability of close-in white dwarf planets in regard to their evolution.  At early times these planets will reach blackbody equilibrium temperatures that exceed 4000~K for short times, which will be sufficient to significantly ablate the rocky surface of a terrestrial planet \citep[e.g.][]{rappaport12,rappaport14,perez-becker}, presenting a challenge for the retention of sufficient water when the WD is cooler and life may develop.

We now consider the limits to such planets around our target stars, as well as two white dwarfs reported in \citet[][hereafter H13]{hoard13} that showed infrared excesses but no clear signature of dust accretion.  These are primary candidates for the phenomenon we consider above.

\subsection{COS Targets}
For each of our COS targets we calculated the black-body equilibrium temperatures for putative irradiated companions in 4 to 30 hour orbits and calculated the limiting radii for the range of orbital periods considered assuming a 3-$\sigma$ detection based on the reported ALLWISE uncertainties (or {\em Spitzer} if contamination of the ALLWISE data was suspected).  We compiled optical photometry from UCAC4 \citep{ucac4}, URAT1\citep{urat1}, or SDSS DR9 \citep{sdss}, near-IR photometry from 2MASS \citep{skrutskie}, and mid-IR photometry from ALLWISE \citep{wright10}  to estimate our sensitivity to irradiated companions.  We also calculated our sensitivity to irradiated companions between 1 and 10~\Rsun away for each target, shown in Figures \ref{fig:irr1} and \ref{fig:irr2}.  For six of the seven WDs, we are sensitive to companions larger than 4~\Rearth, and for a subset we are sensitive down to 2~\Rearth.  We note any special cases below.

{\bf WD 0308-565:}  This target is unfortunately confused in the WISE channels. Higher resolution {\em Spitzer} IRAC2 and IRAC4 images clearly show another source close to WD 0308-565's position.  Both Spitzer channels are consistent within the flux calibration uncertainties to the expected photospheric values.  We thus use the IRAC2 channel flux (assuming 5\% absolute uncertainties in the photometry) to constrain the presence of any irradiated companions, using the reported fluxes in the 3\farcs8 aperture in the SEIP catalog \citep{seip}.  Based on second epoch observations, the secondary point source is not associated with WD 0308-565.

{\bf GD~153:} Unlike the other WDs, we could not retrieve reliable optical photometry.  We therefore relied on the values reported in \citet{landolt13} and assumed 5\% uncertainties.

{\bf WD 1057+719:}  Like WD 0308-565, W1 and W2 channels are contaminated by a visual companion discovered in higher resolution {\em Spitzer} IRAC images.  The reported photometry from the SEIP catalog is consistent with pure photospheric emission, and we used the upper limits to excesses at all four IRAC channels to determine the limiting radius to which we were sensitive with this target.  For this target the 8\micron\ images are the most sensitive to close-in companions.

{\bf SDSSJ1228}:  Since this WD has a strong IR excess, it is not possible to detect the signature of a small planet in a close orbit.

\subsection{H13 Excess Candidates}
We consider three WDs reported in H13; WD 0249-052, WD 1046-017, and WD 1448+411.  All three of these WDs were shown to have IR excesses in the WISE All-Sky Catalog, yet two of the three (WD 0249-052, DB; WD 1046-017, DB) had no published metal line detections with high resolution echelle spectra, despite being DBs with sensitive upper limits to accretion.  WD 1448+411 has since been shown to possess no clear metal line signatures (S. Xu 2013, private comm.) as well. The lack of metal lines of a given equivalent width in hot white dwarfs can be due to the fact that the Ca II H and K lines, the primary atomic species for detecting accretion, become weaker at higher T$_{eff}$ and can be invisible for low accretion rates \citep{debes11}.  Conversely, a substellar companion would cause the infrared excess but would not be expected to put material onto the WD surface unless it was actively transferring mass \citep{xu15}.   While H13 listed WD 1046-017 as a candidate based on its WISE All-Sky Catalog photometry, the excess has since been shown to be a spurious detection, because revised photometry reported in the ALLWISE Reject Catalog show that WD 1046-017's W1 photometry is consistent with the WD photosphere.  WD 0249-052's WISE photometry from the ALLWISE catalog revises the excess (as presented in Figure \ref{fig:024}) smaller but still tentatively present at 2.4 and 2.8~$\sigma$ above the predicted WD photosphere.  The excess observed for WD~1448+411 remains securely detected with new ALLWISE photometry.  For both of these WDs, we consider objects orbiting just outside the tidal disruption radius of the WD, but with equilibrium temperatures that are still in excess of 1000~K.

WD 0249-052's candidate excess is quite faint, and at the limit of what may be detected with ALLWISE photometry.  As with all WISE selected candidates, the possibility for contamination is present, but DSS and 2MASS images show no obvious blends.  Figure \ref{fig:024} shows the SED of this WD along with its excess.  Both a T0.5 unresolved companion taken from empirical SEDs of field brown dwarfs \citep{debes11} or an irradiated blackbody with a similar T$_{eff}$ of 1400~K and a radius of 5.2\Rearth\ provide a reasonable fit to the photometry without the need for a disk of dust.  Since this companion would be in the L/T transition if a brown dwarf, it may also show variability in the near-IR \citep{radigan12}. 

WD 1448+411 has brighter emission, and is comparable to an L8.5 companion when compared to empirical SEDs of field brown dwarfs.  When we consider an irradiated planet (See Figure \ref{fig:144}), we find that a T$_{eff}$=1140~K 0.9 R$_{\rm J}$ planet could also explain the excess.  WD~1448+411 lacks published NIR photometry or spectroscopy which potentially could differentiate between these two possibilities--the cooler irradiated body would have no significant excess below $\sim$3\micron.

\section{Discussion}
\label{sec:disc}
We have demonstrated that as a high-speed photometer, COS is photon-limited for short time exposures with high countrates, and is sensitive to small levels of variability from objects with UV flux.  In determining the sensitivity of the FUV detector on COS, we observed long-term 2\% amplitude photometric fluctuations in the lightcurve of GD 133, with a period of 5.2 hrs.  In our study of seven hot WDs, we were able to look for small deviations in the light curves that could potentially indicate an astrophysical event.  We found that we were capable of detecting transiting sub-Earth objects around each WD, even with our faintest objects.  We also discovered that we are capable of detecting heavily irradiated super-Earth planets in close-in orbits around hot ($>$15,000 K) WDs, and identified two potential candidates that show an infrared excess without evidence of metal accretion.

With the level of sensitivity we were able to achieve, we could theoretically observe large transiting planetesimals around a WD with a debris disk.  The planetesimals would most likely be just outside of the tidal disruption region of the WD, and in highly eccentric orbits making it much harder to pin down a certain orbital period.  A good example is the recent discovery of evaporating planetesimals in orbit around WD 1145+017 \citep{vanderberg15,croll15,xu15}.  In addition to this possibility, COS could also be used to look for short-term variability in the absorption features of A-stars that could indicate the presence of Òfalling evaporated bodiesÓ (FEBs) from their circumstellar disks \citep{welsh13}. 

COS could also utilize its spectroscopic capabilities to investigate the atmospheric composition of any transiting planet around a WD by measuring its effective radius as a function of wavelength.  Both molecular oxygen (O$_{2}$) and ozone (O$_{3}$) have significant opacity in the UV.  The larger $O_{2}$ signal peaks in the FUV, while the smaller $O_{3}$ signal peaks in the NUV \citep{betremieux13}.  A large signal of molecular oxygen could indicate the presence of various biological processes (e.g. photosynthesis), and thus be a bio-marker in conjunction with O$_3$ and CH$_4$; although this is dependent on the specifics of the input flux to the planet \citep{dg14}.  In any case, terrestrial planets, if detected around a sufficiently bright WD, could be accessible to direct atmospheric characterization years before it could be done for a planet around an M-dwarf or solar-type star.  While WDs are unlikely hosts for close-in habitable planets due to their evolutionary history, they do represent a useful population that is uniquely well suited for study with existing technologies and for a modest investment of telescope time.

\acknowledgements
PHS and JHD thank the Baltimore Ingenuity Project (www.ingenuityproject.org) for funding and support for this research project.  Support for this work was provided by NASA through grants HST-GO-13752 and HST-AR-13902, and  from the Space Telescope Science Institute, which is operated by AURA, Inc., under NASA contract NAS 5-26555.  The authors wish to thank Matt Lallo for helpful discussions about behavior of HST with regards to focus, Steve Penton for investigating data anomalies we discovered as part of our work, and Peter McCullough for helpful discussions of transit detection.  The authors also wish to thank Pier-Emannuel Tremblay for graciously providing model WD spectra for our ETC calculations.  This publication makes use of data products from the Two Micron All Sky Survey, which is a joint project of the University of Massachusetts and the Infrared Processing and Analysis Center/California Institute of Technology, funded by the National Aeronautics and Space Administration and the National Science Foundation, from GALEX, a NASA mission managed by the Jet Propulsion Laboratory in partnership with California Institute of Technology, from the Wide-field Infrared Survey Explorer, which is a joint project of the University of California, Los Angeles, and the Jet Propulsion Laboratory/California Institute of Technology, and from NEOWISE, which is a project of the Jet Propulsion Laboratory/California Institute of Technology. WISE and NEOWISE are funded by the National Aeronautics and Space Administration.  This research has made use of the VizieR catalogue access tool and the SIMBAD database, CDS,
 Strasbourg, France. The original description of the VizieR service was published in A\&AS 143, 23.

\begin{deluxetable}{ccccccccc}
\tablecolumns{9}
\tablewidth{0pt}
  \tablecaption{Limiting V magnitude for COS WD Lightcurves, S/N=10, 30~s sampling \label{tab:etc}}
\tablehead{
\colhead{T$_{\rm eff}$} & \colhead{G130M} & \colhead{G160M} & \colhead{G140L} & \colhead{G185M} & \colhead{G225M} & \colhead{G285M} & \colhead{G230L} & \colhead{NUV} \\
\colhead{K} & \colhead{1309\AA} & \colhead{1600\AA} & \colhead{1280\AA} & \colhead{1850\AA} & \colhead{2250\AA} & \colhead{2850} & \colhead{2950} & }
\startdata
5000 & ... & ... & ... & ... & ... & ... & ... & 16.0 \\
10000 & 12.2 & 15.0 & 15.4 & 15.2 & 15.7 & 15.0 & 18.0 & 22.0 \\
15000 & 17.5 & 17.9 & 17.9 & 15.8 & 15.6 & 14.0 & 17.9 & 20.5 \\
20000 & 22.0 & 21.5 & 21.9 & 18.9 & 18.1 & 16.0 & 20.6 & 23.0 \\
\enddata
\end{deluxetable}

\begin{deluxetable}{cccccccc}
\tablecolumns{8}
\tablewidth{0pt}
  \tablecaption{Properties of Transit Search Targets \label{tab:targets}}
\tablehead{
\colhead{Name} & \colhead{T$_{\rm eff}$} & \colhead{$\log$~g} & \colhead{$\log$ L$_{\rm WD}$} & \colhead{M$_{\rm WD}$} & \colhead{R$_{\rm WD}$} & \colhead{R$_{\rm PHZ,i}$} & \colhead{R$_{\rm PHZ,o}$}  \\
 & \colhead{K} &  & \colhead{L$_\odot$} & \colhead{M$_\odot$} & \colhead{R$_\odot$} & \colhead{AU} & \colhead{AU} }
\startdata
WD 0308-565 & 22849 & 8.06 & -1.4 & 0.64 & 0.012 & 0.14 & 0.39 \\
G 191-B2b & 60920 & 7.55 & 0.71 & 0.54  & 0.020 & 0.64 & 3.4 \\
GD 71 & 33590 & 7.93 & -0.64 & 0.62 & 0.014 & 0.35 & 0.73 \\
GD 153 & 40320 & 7.93 & -0.31 & 0.64 & 0.014 & 0.50 & 1.07 \\
WD 0947+857 & 50890 & 8.23 & -0.10 & 0.82 & 0.011 & 0.64 & 1.36 \\
WD 1057+719 & 42050 & 7.85 & -0.18 & 0.60 & 0.015 & 0.59 & 1.24 \\
SDSS J1228+1040 & 20900 & 8.15 & -1.64 & 0.71 & 0.012 & 0.11 & 0.23 \\
\enddata
\end{deluxetable}
\appendix

\section{Short Exposure COS observations}
\label{app:A}
In Table \ref{atab:cos} we list a sample of the COS observations of WDs used to create Figure \ref{fig:cos}.   A machine readable table is available in the online version of this manuscript.

 \begin{deluxetable}{ccccccc}
 \tablecolumns{7}
\tablewidth{0pt}
 \tabletypesize{\scriptsize}
 \tablecaption{\label{atab:cos} Short Exposure COS observations}
\tablehead{\colhead{Name} &  \colhead{Exptime} & \colhead{Count Rate} & \colhead{Grating} & \colhead{CENWAVE} & \colhead{Proposal ID} & \colhead{File Name}}
\startdata
SDSSJ084539.17+225728.0  & 510 & 18858 & G130M & 1291 & 11561 & lb5z01niq\_corrtag\_a.fits \\
SDSSJ084539.17+225728.0 & 510 & 18641 & G130M & 1291 & 11561 & lb5z01nkq\_corrtag\_a.fits \\
SDSSJ084539.17+225728.0 & 465 & 19456 & G130M & 1327 & 11561 & lb5z01nmq\_corrtag\_a.fits \\
SDSSJ084539.17+225728.0  & 555 & 19581 & G130M & 1327 & 11561 & lb5z01noq\_corrtag\_a.fits \\
SDSSJ084539.17+225728.0  & 1485 & 15275 & G160M & 1577 & 11561 & lb5z01nqq\_corrtag\_a.fits \\
SDSSJ084539.17+225728.0  & 2235 & 12329 & G160M & 1623 & 11561 & lb5z01nsq\_corrtag\_a.fits \\
SDSSJ122859.93+104032.9 & 525 & 25343 & G130M & 1291 & 11561 & lb5z03feq\_corrtag\_a.fits \\
\enddata
\end{deluxetable}

\section{Transit Survey COS observations}
\label{app:B}
We list here sample tables of COS observations for the WDs surveyed for transits in \S \ref{sec:search}.  Full machine readable versions of the tables are available in the online version of this manuscript.

\begin{deluxetable}{ccccccc}
 \tablecolumns{7}
 \tabletypesize{\scriptsize}
\tablewidth{0pt}
 \tablecaption{COS observations of WD 0308-565}

\tablehead{ \colhead{Start Date} & \colhead{Exptime} & \colhead{Count Rate} & \colhead{Grating} & \colhead{CENWAVE} & \colhead{Proposal D} & \colhead{File}}
\startdata
 2011-01-07 05:48:50.301 & 480 & 122563 & G130M & 1300 & 12426 & lbnm02a2q\_corrtag\_a.fits \\
2011-01-07 05:59:46.301 & 480 & 121215 & G130M & 1300 & 12426 & lbnm02a4q\_corrtag\_a.fits \\
 2011-01-07 06:10:42.333 & 480 & 120807 & G130M & 1300 & 12426 & lbnm02a6q\_corrtag\_a.fits \\
2011-01-07 06:50:22.333 & 480 & 125232 & G130M & 1309 & 12426 & lbnm02arq\_corrtag\_a.fits \\
\enddata
\end{deluxetable}

\begin{deluxetable}{cccccccc}
 \tablecolumns{7}
 \tabletypesize{\scriptsize}
\tablewidth{0pt}
 \tablecaption{COS observations of GD 71}

\tablehead{ \colhead{Start\_Date} & \colhead{Exptime} & \colhead{Count Rate} & \colhead{Grating} & \colhead{CENWAVE} & \colhead{Proposal\_ID} & \colhead{File}}
\startdata
 2009-09-22 01:00:05.062 & 60 & 170577 & G230L & 2635 & 11481 & la9g02jvq\_corrtag.fits \\
 2009-09-22 01:03:14.150 & 60 & 163740 & G230L & 2635 & 11481 & la9g02jyq\_corrtag.fits \\
 2009-09-22 01:06:23.046 & 60 & 157291 & G230L & 2635 & 11481 & la9g02k0q\_corrtag.fits \\
2009-09-22 01:09:32.133 & 60 & 151664 & G230L & 2635 & 11481 & la9g02k2q\_corrtag.fits \\
\enddata
\end{deluxetable}

\begin{deluxetable}{cccccccc}
 \tablecolumns{7}
 \tabletypesize{\scriptsize}
\tablewidth{0pt}
 \tablecaption{COS observations of G 191-B2b}

\tablehead{ \colhead{Start\_Date} & \colhead{Exptime} & \colhead{Count Rate} & \colhead{Grating} & \colhead{CENWAVE} & \colhead{Proposal\_ID} & \colhead{File}}
\startdata
 2009-09-26 08:21:34.054 & 60 & 153442 & G225M & 2186 & 11481 & la9g01luq\_corrtag.fits \\
 2009-09-26 08:24:58.054 & 60 & 153300 & G225M & 2186 & 11481 & la9g01lwq\_corrtag.fits \\
 2009-09-26 08:28:22.054 & 60 & 152492 & G225M & 2186 & 11481 & la9g01lyq\_corrtag.fits \\
2009-09-26 08:31:46.054 & 60 & 152277 & G225M & 2186 & 11481 & la9g01m0q\_corrtag.fits \\
\enddata
\end{deluxetable}

\begin{deluxetable}{cccccccc}
 \tablecolumns{7}
 \tabletypesize{\scriptsize}
\tablewidth{0pt}
 \tablecaption{COS observations of GD 153}

\tablehead{ \colhead{Start\_Date} & \colhead{Exptime} & \colhead{Count Rate} & \colhead{Grating} & \colhead{CENWAVE} & \colhead{Proposal\_ID} & \colhead{File}}
\startdata
 2010-04-30 22:04:25.532 & 45 & 133039 & G230L & 2635 & 11896 & lbbd10djq\_corrtag.fits \\
2010-04-30 22:08:21.532 & 135 & 226537 & G230L & 2950 & 11896 & lbbd10dlq\_corrtag.fits \\
2010-04-30 23:19:43.516 & 1305 & 175227 & G230L & 3360 & 11896 & lbbd10dnq\_corrtag.fits \\
2010-05-31 03:08:30.524 & 45 & 133086 & G230L & 2635 & 11896 & lbbd11alq\_corrtag.fits \\

\enddata
\end{deluxetable}

\begin{deluxetable}{cccccccc}
 \tablecolumns{7}
 \tabletypesize{\scriptsize}
\tablewidth{0pt}
 \tablecaption{COS observations of WD 0947+857}
\tablehead{ \colhead{Start\_Date} & \colhead{Exptime} & \colhead{Count Rate} & \colhead{Grating} & \colhead{CENWAVE} & \colhead{Proposal\_ID} & \colhead{File}}
\startdata
 2009-09-14 07:38:01.350 & 450 & 92535 & G140L & 1230 & 11494 & la9r01cdq\_corrtag\_a.fits \\
 2009-09-14 08:30:29.318 & 450 & 90593 & G140L & 1230 & 11494 & la9r01ciq\_corrtag\_a.fits \\
 2009-09-14 08:40:40.325 & 450 & 93108 & G140L & 1230 & 11494 & la9r01ckq\_corrtag\_a.fits \\
 2009-09-14 08:51:58.342 & 720 & 92520 & G140L & 1230 & 11494 & la9r01cmq\_corrtag\_a.fits \\
\enddata
\end{deluxetable}

\begin{deluxetable}{cccccccc}
 \tablecolumns{7}
 \tabletypesize{\scriptsize}
\tablewidth{0pt}
 \tablecaption{COS observations of WD 1057+719}
\tablehead{ \colhead{Start\_Date} & \colhead{Exptime} & \colhead{Count Rate} & \colhead{Grating} & \colhead{CENWAVE} & \colhead{Proposal\_ID} & \colhead{File}}
\startdata
2009-09-14 14:53:05.318 & 330 & 152646 & G160M & 1600 & 11494 & la9r02dhq\_corrtag\_a.fits \\
2009-09-14 15:01:40.326 & 330 & 155627 & G160M & 1600 & 11494 & la9r02djq\_corrtag\_a.fits \\
2009-09-14 15:10:15.334 & 330 & 157986 & G160M & 1600 & 11494 & la9r02dlq\_corrtag\_a.fits \\
 2009-09-14 15:18:50.342 & 330 & 161098 & G160M & 1600 & 11494 & la9r02dnq\_corrtag\_a.fits \\
\enddata
\end{deluxetable}

\begin{deluxetable}{cccccccc}
 \tablecolumns{7}
 \tabletypesize{\scriptsize}
\tablewidth{0pt}
 \tablecaption{COS observations of SDSS J1228+1040}
\tablehead{ \colhead{Start\_Date} & \colhead{Exptime} & \colhead{Count Rate} & \colhead{Grating} & \colhead{CENWAVE} & \colhead{Proposal\_ID} & \colhead{File}}
\startdata
 2010-04-12 15:40:54.811 & 690 & 19003 & G130M & 1291 & 11561 & lb5z03feq\_corrtag\_a.fits \\
 2010-04-12 15:54:52.796 & 690 & 18981 & G130M & 1291 & 11561 & lb5z03fjq\_corrtag\_a.fits \\
 2010-04-12 16:09:59.804 & 600 & 18823 & G130M & 1327 & 11561 & lb5z03flq\_corrtag\_a.fits \\
 2010-04-12 17:06:11.804 & 810 & 18808 & G130M & 1327 & 11561 & lb5z03fvq\_corrtag\_a.fits \\
\enddata
\end{deluxetable}

\clearpage

\bibliography{wd_chap}
\bibliographystyle{apj}

\begin{figure}
\plotone{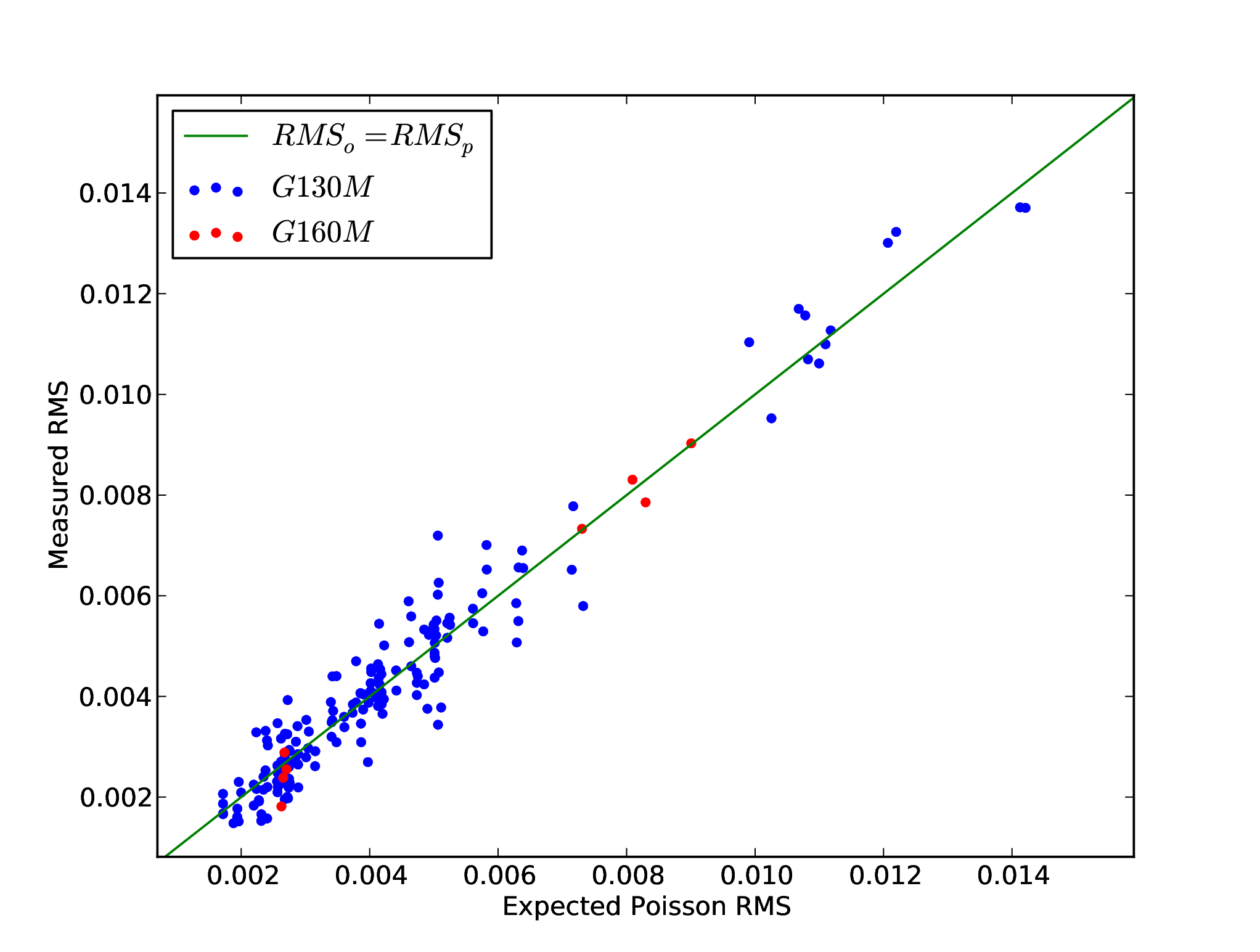}
\caption{\label{fig:cos} We measure the fractional RMS of count-rates within 20s bins for single exposures of WDs with COS and compare them to the expected fractional RMS of count-rates assuming Poisson counting statistics.  The blue dots are for targets with the G130M FUV grating.  This grating corresponds to wavelengths from 1150 to 1450 \AA.  The red dots represent targets that fall into the category of the G160M FUV grating, with coverage from 1405 to 1775 \AA.   The green line shows a graph of the line RMS$_{\rm obs}$=RMS$_{pred}$.  This line indicates that the precision of COS photometry at moderate to high count rates is only limited by the number of photons detected.}
\end{figure}

\begin{figure}
\plotone{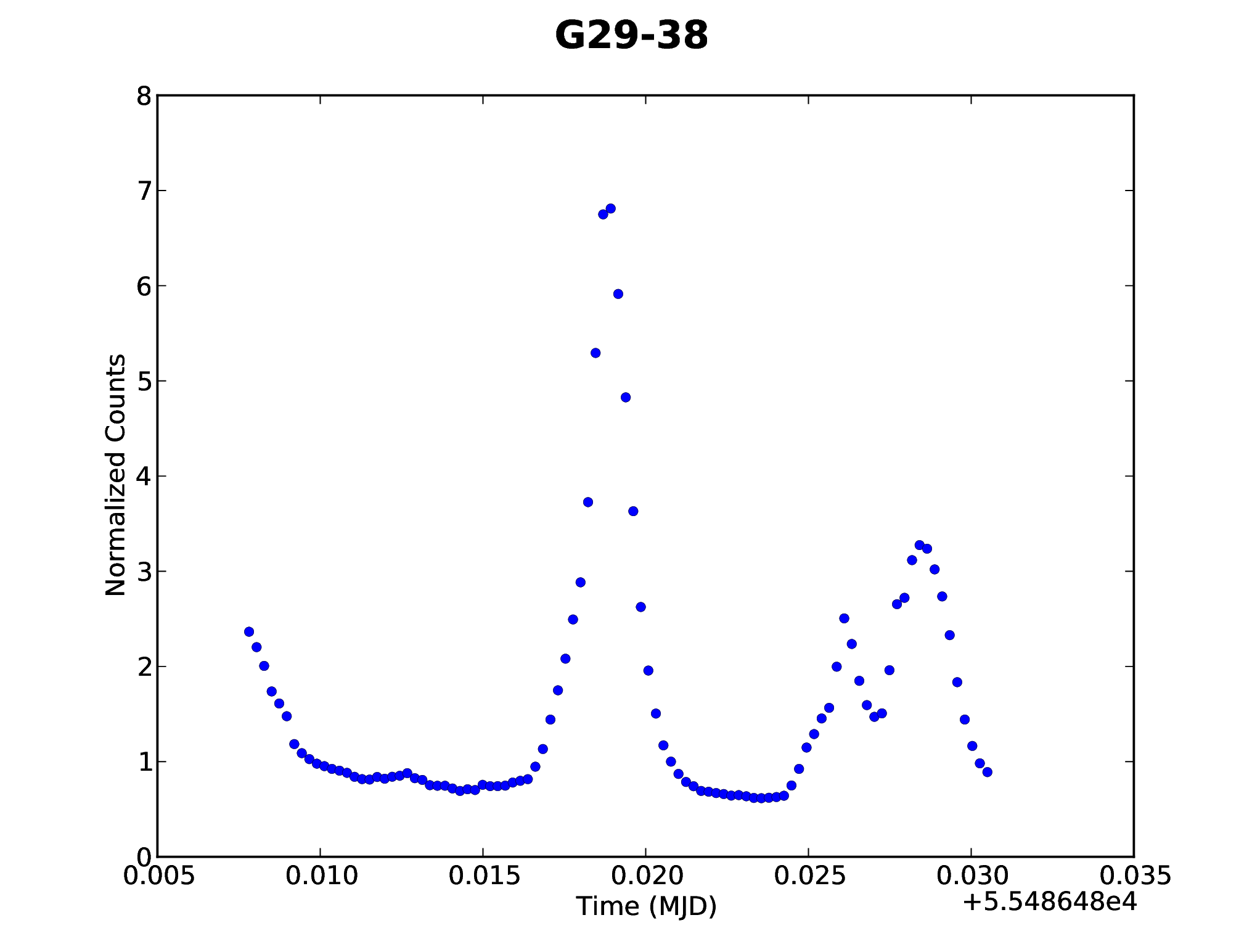}
\caption{\label{fig:G29-38} One of four observations of G29-38, showing the short period, large amplitude pulsations.  The fluctuations occur every ~8-11 min, and differ greatly in magnitude.}
\end{figure}

\begin{figure}
\plotone{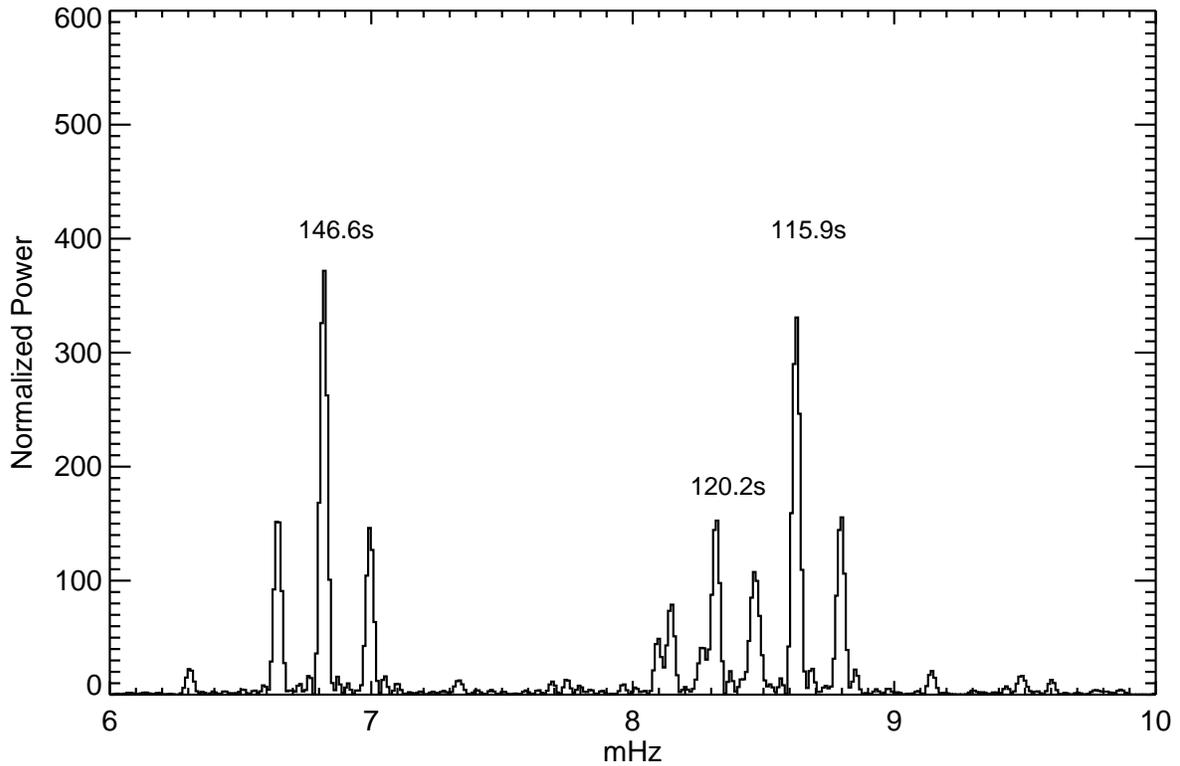}
\caption{\label{fig:lomb} Zoom in of the 6-10mHz region of GD~133's Lomb-Scargle normalized periodogram, showing strong peaks at periods similar to those seen in optical high speed photometry of the WD.  The amplitudes of the 146.6~s and 115.9~s pulsations are 25~mma and 24~mma respectively.  The observed side-lobes are due to the 50-minute orbital visibility windows of HST.}
\end{figure}

\begin{figure}
\plotone{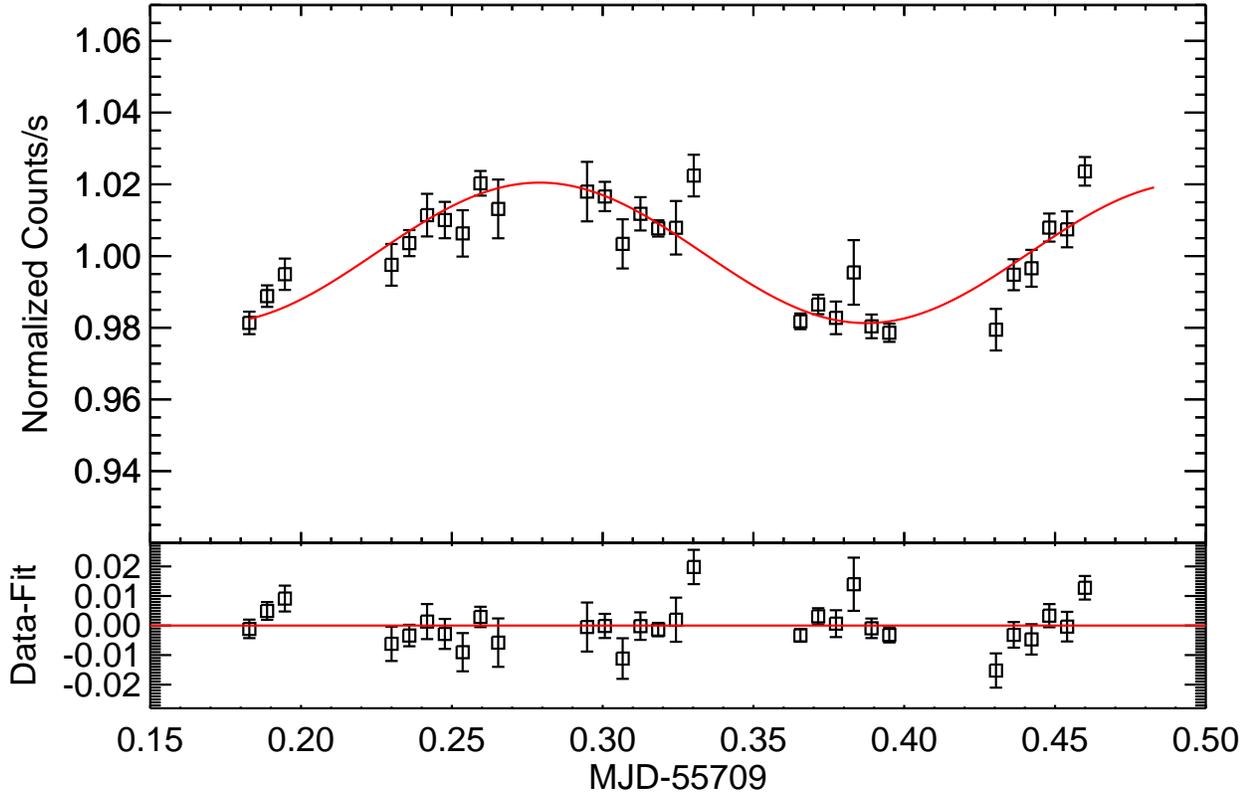}
\caption{\label{fig:long} Long period variability in GD~133, showing a 5.2~hr period and 1.9\% amplitude.  The observed countrates were corrected both for varying position of the spectrum on the detector and the changing focus of HST.}
\end{figure}

\begin{figure}
\plottwo{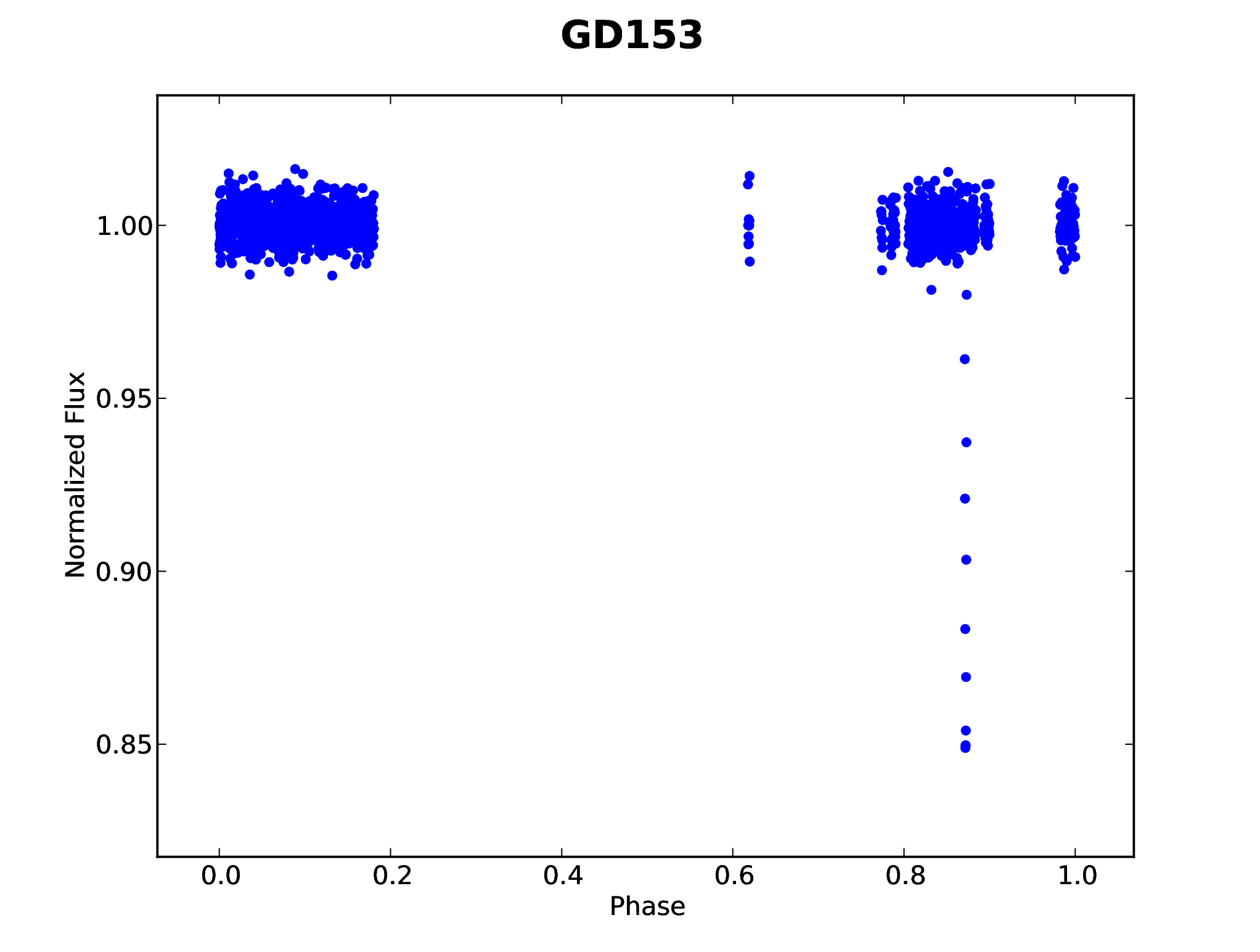}{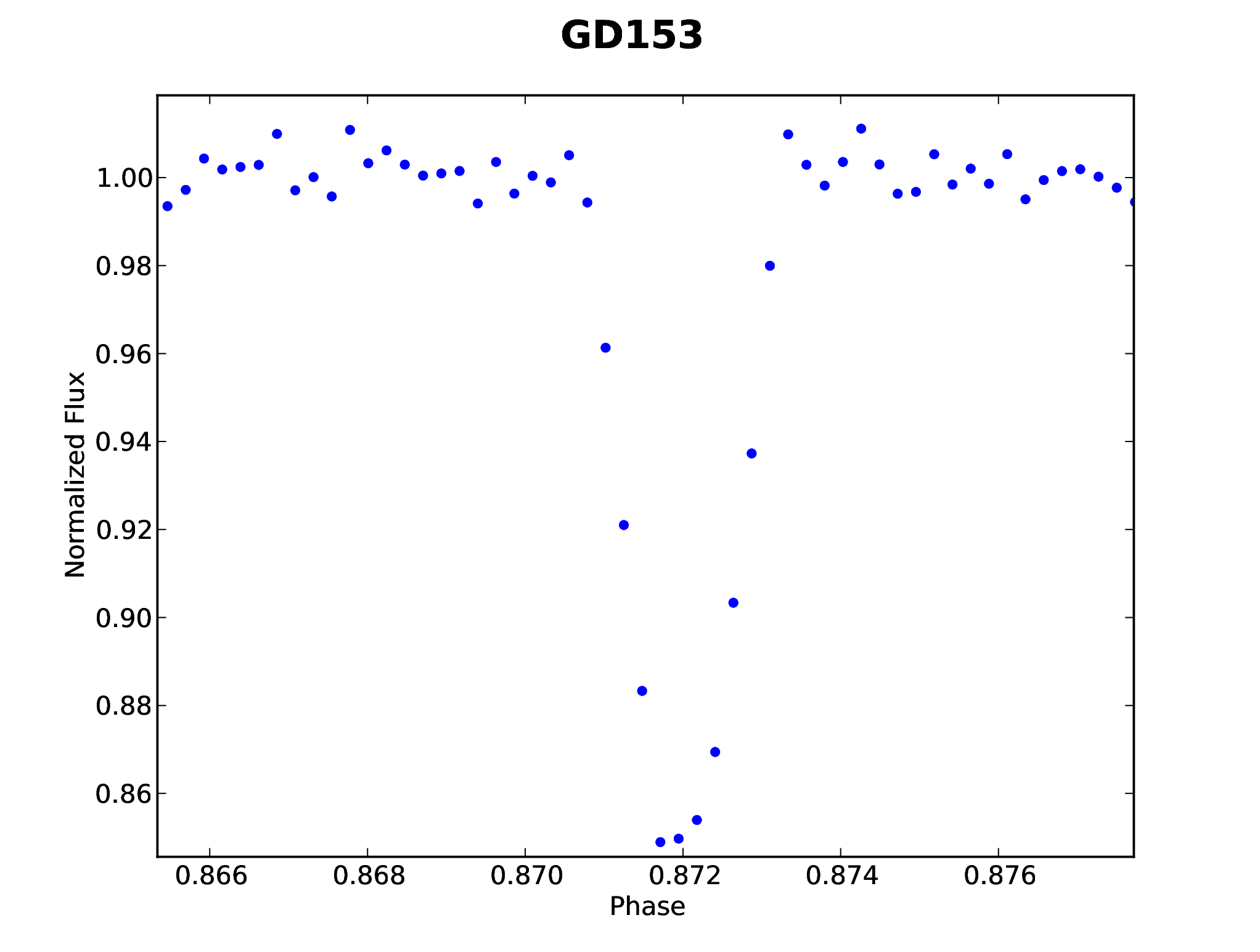}
\caption{\label{fig:tran} Artificial transit for planet with a radius of 0.5 {\Rearth} inserted into GD153 data folded on a 6 hr period.}
\end{figure}

\begin{figure}
\plottwo{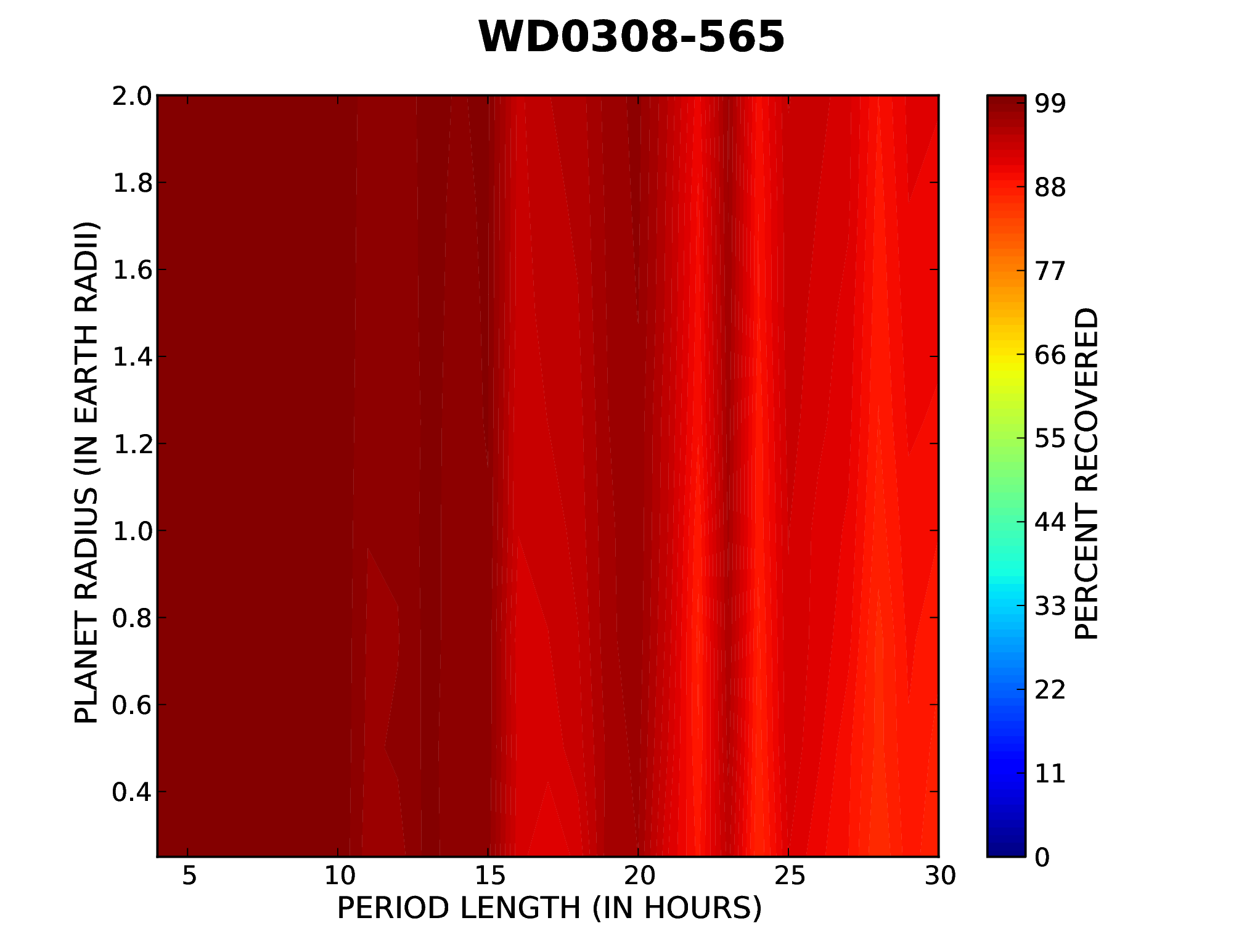}{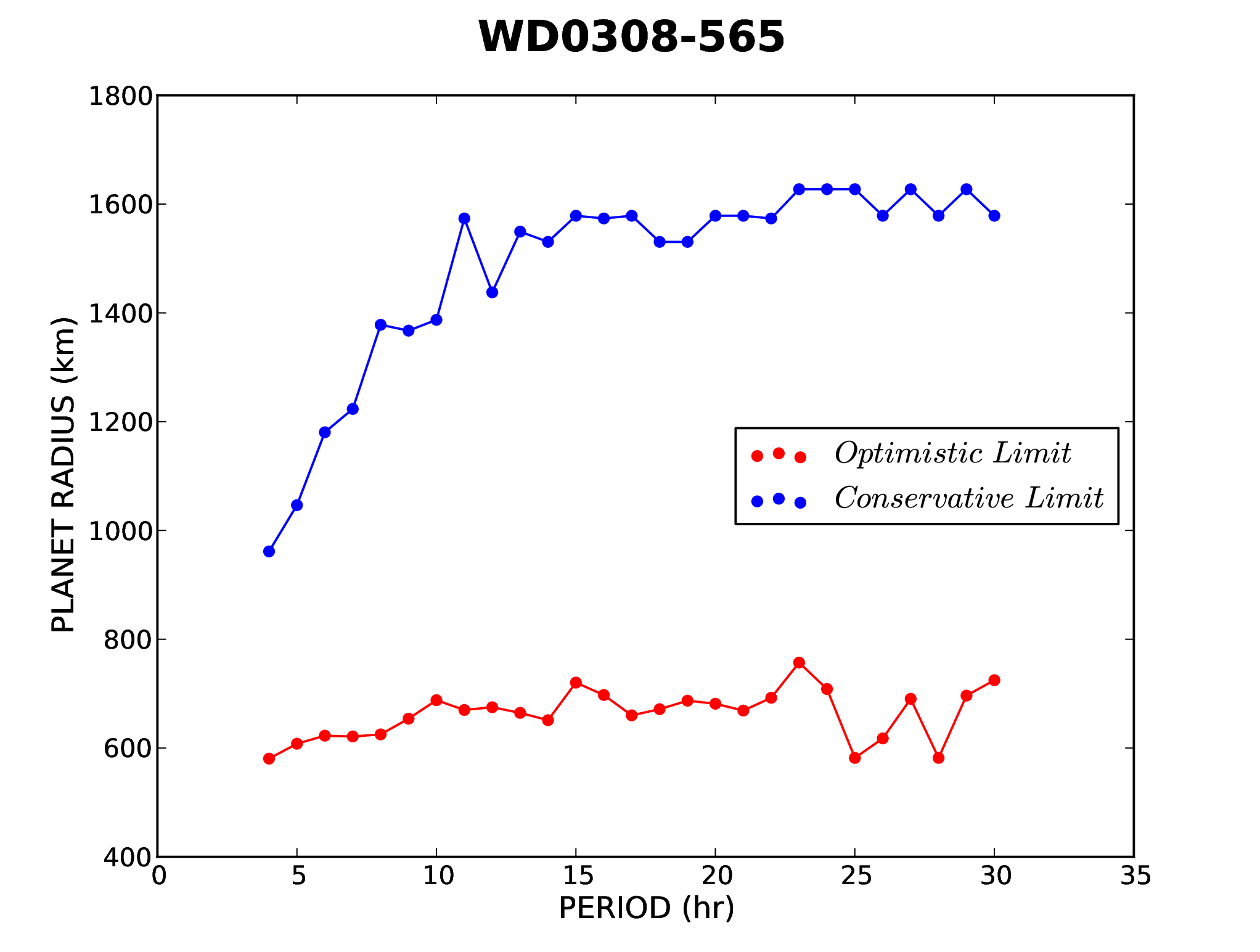}
\caption{\label{fig:0308} (Left) We present detection limits of transiting planets for WD 0308-565.  The contour plot shows the percentage of artificial injected transits recovered in WD 0308-565's lightcurve as a function of planet radius and orbital period. The recovery rates are roughly insensitive to the radius of the companion, provided that its transit non-grazing and the detection is $>5~\sigma$. (Right) The optimistic (red) and conservative (blue) upper limits in radius to possible transiting companions for WD 0308-565 data folded on a given period and rebinned.  These limits are assumed to have a similar recovery rate as in the left panel.  See Section \ref{sec:wdlc:analysis} for more discussion.}
\end{figure}

\begin{figure}
\plottwo{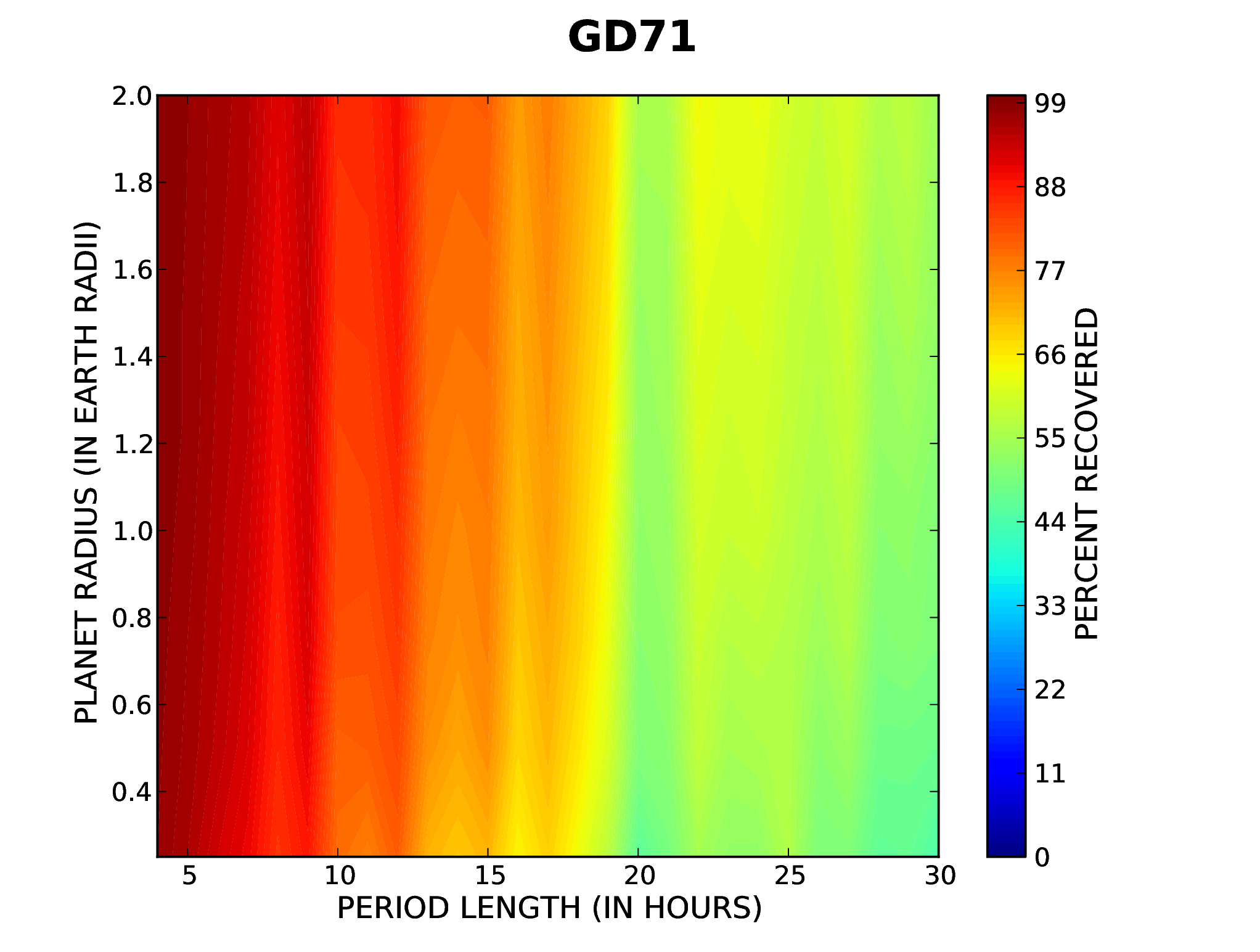}{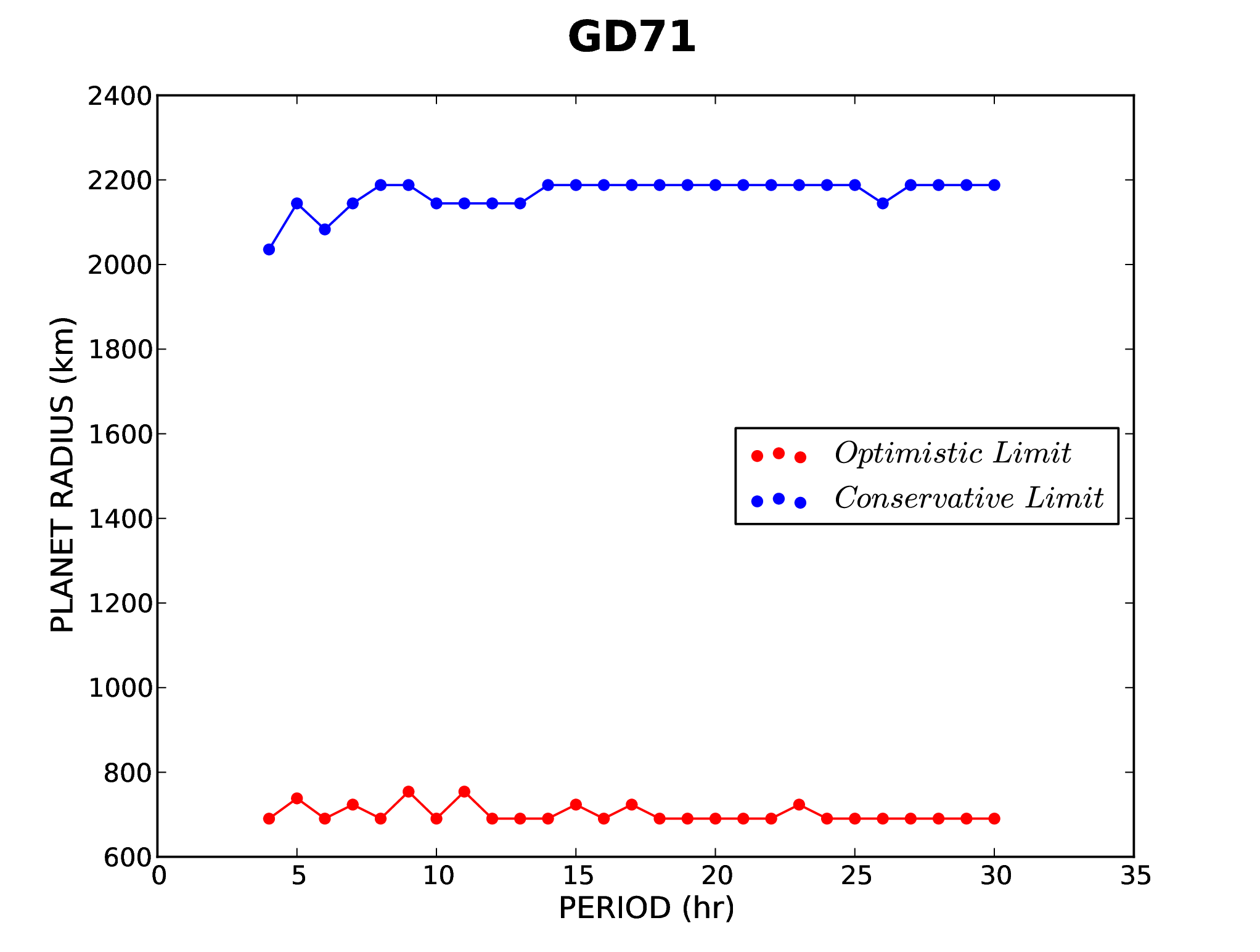}
\caption{\label{fig:71} Same as Figure \ref{fig:0308} for GD71.}
\end{figure}

\begin{figure}
\plottwo{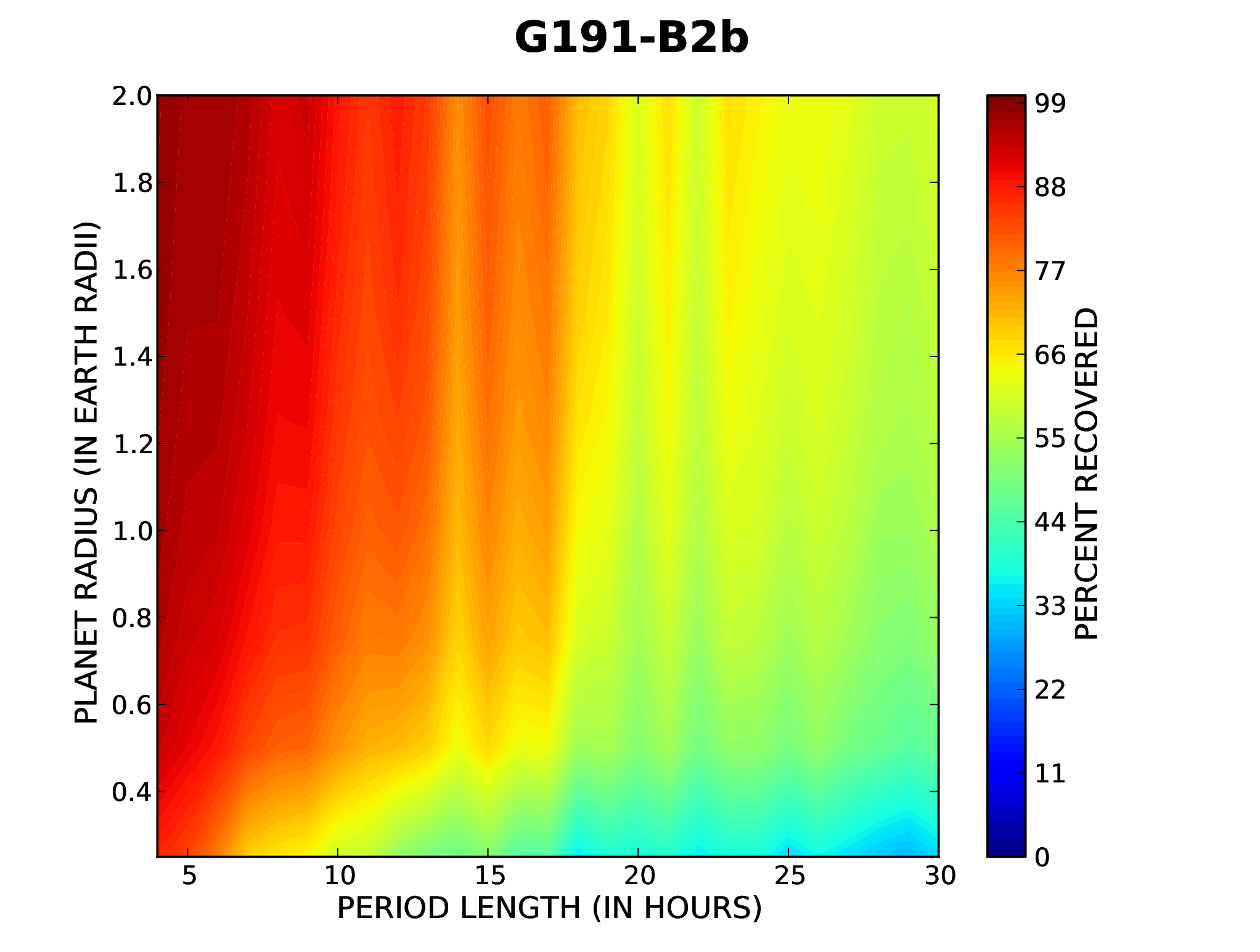}{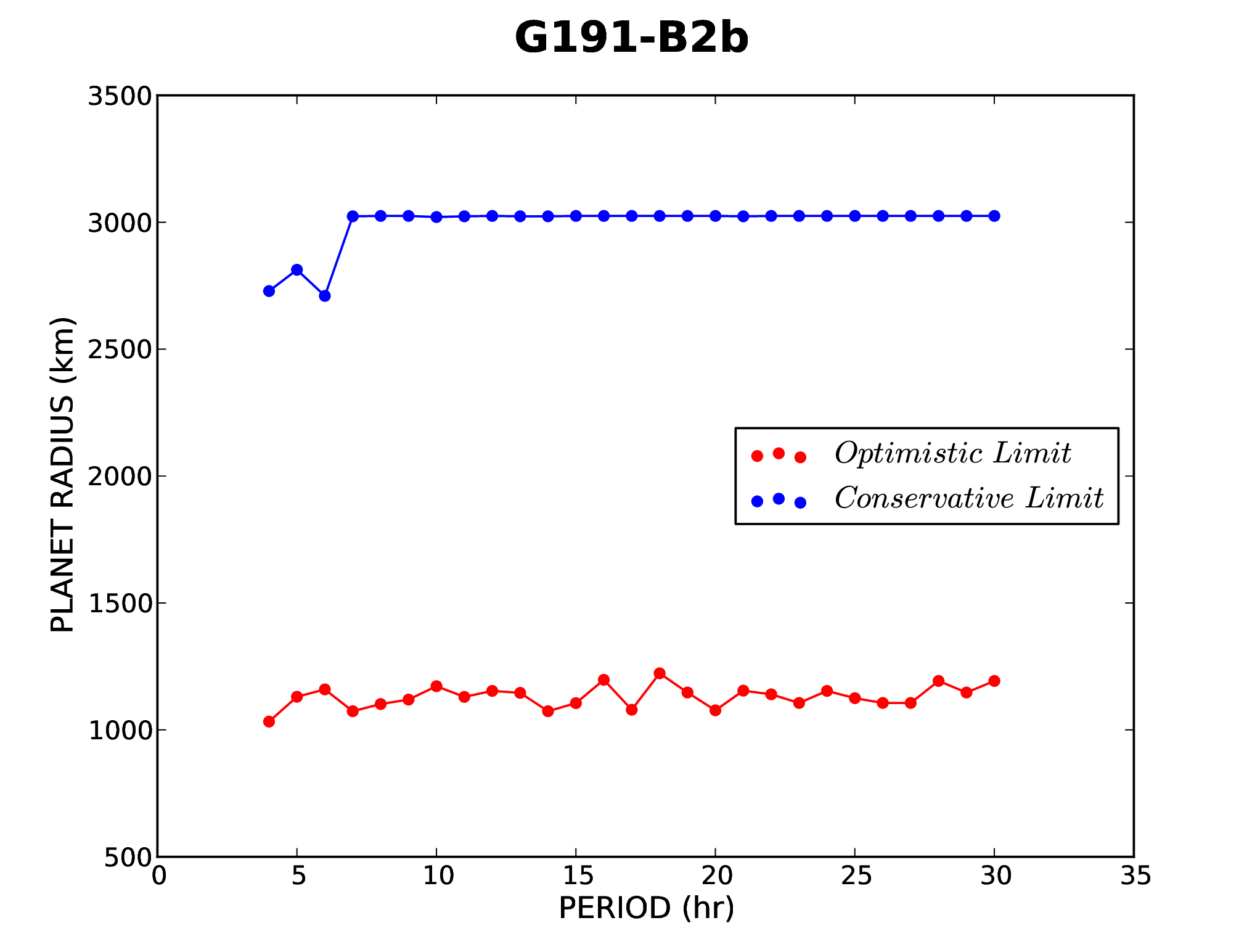}
\caption{\label{fig:191-B} Same as Figure \ref{fig:0308} for G191-B2b.}
\end{figure}

\begin{figure}
\plottwo{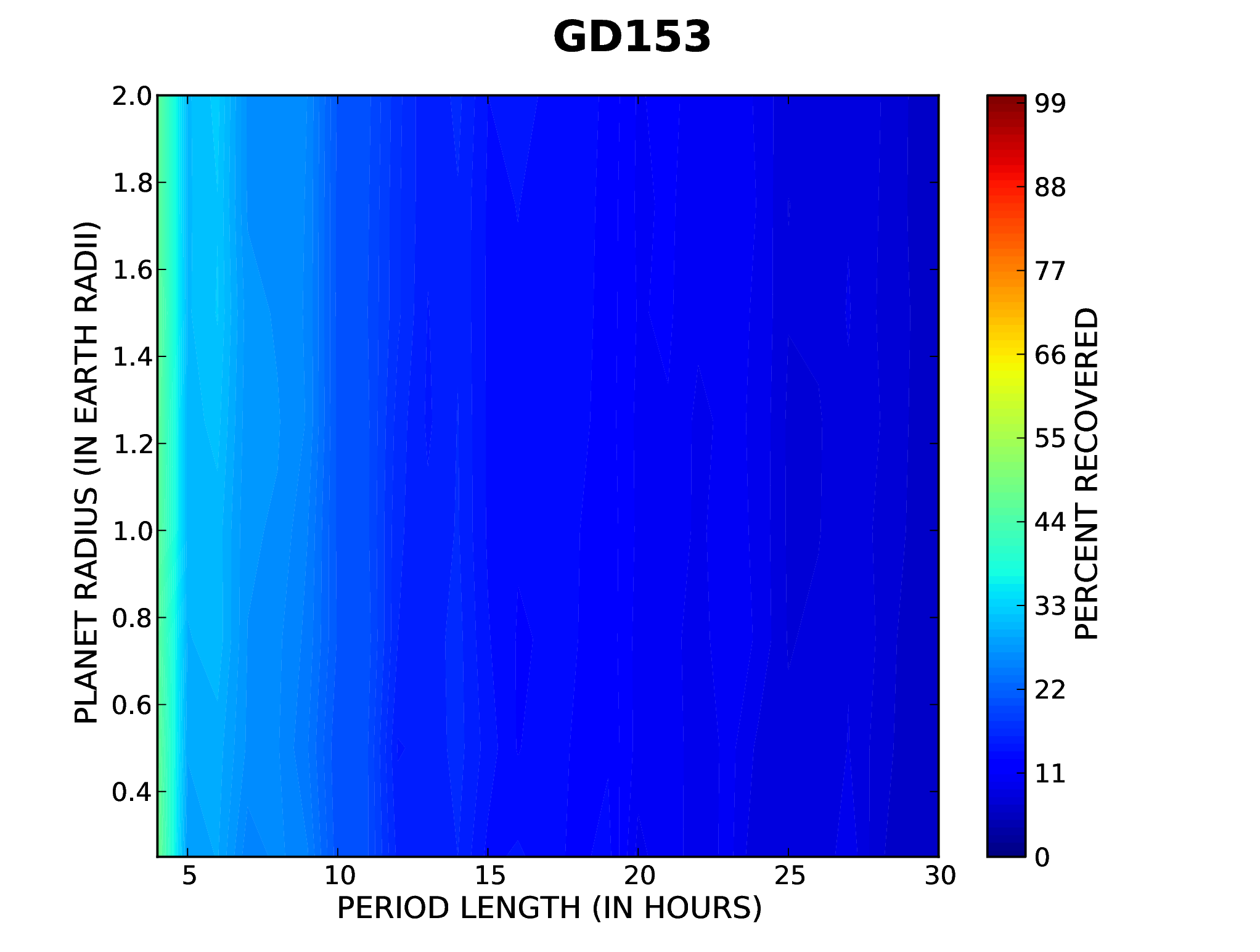}{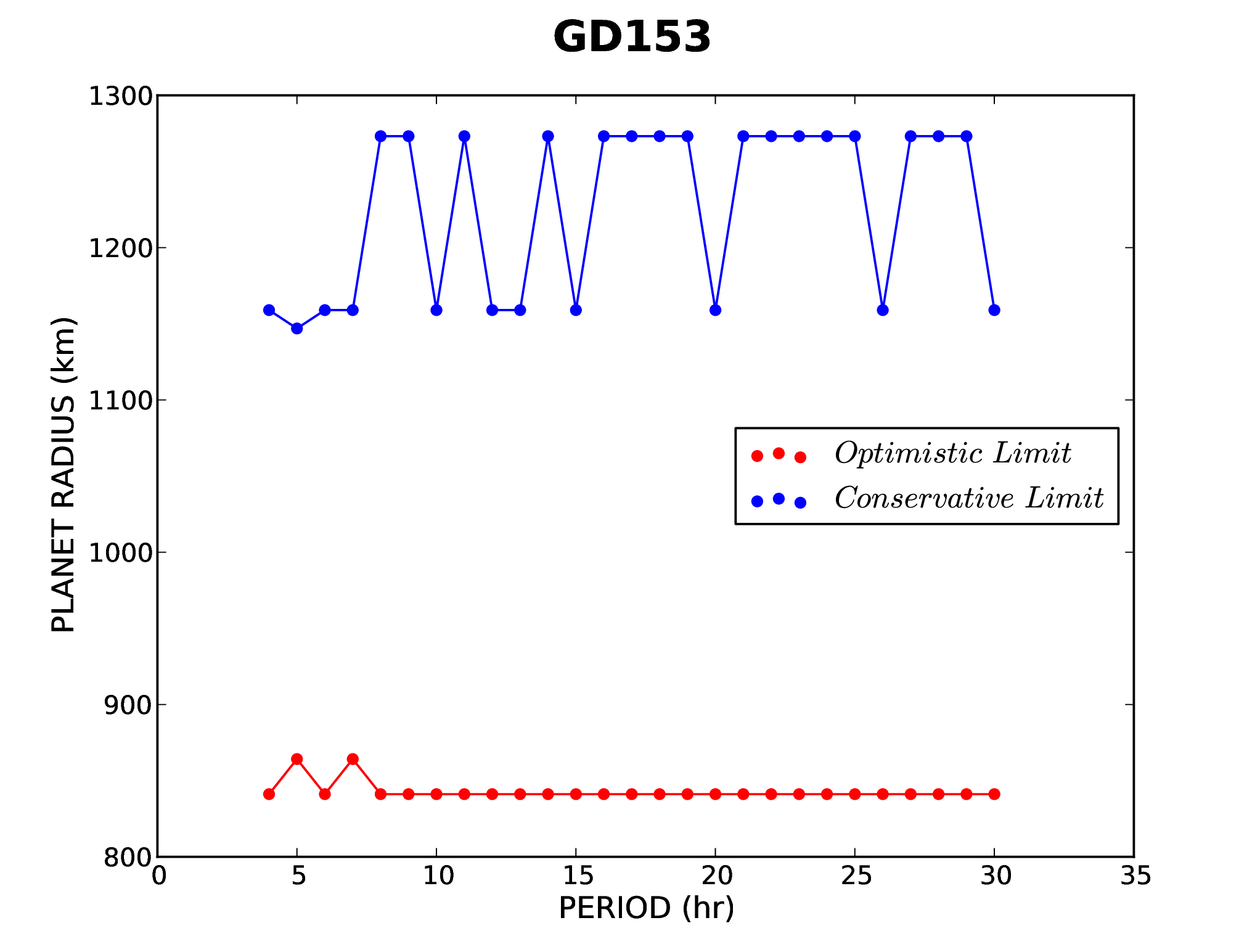}
\caption{\label{fig:GD153} Same as Figure \ref{fig:0308} for GD153.}
\end{figure}

\begin{figure}
\plottwo{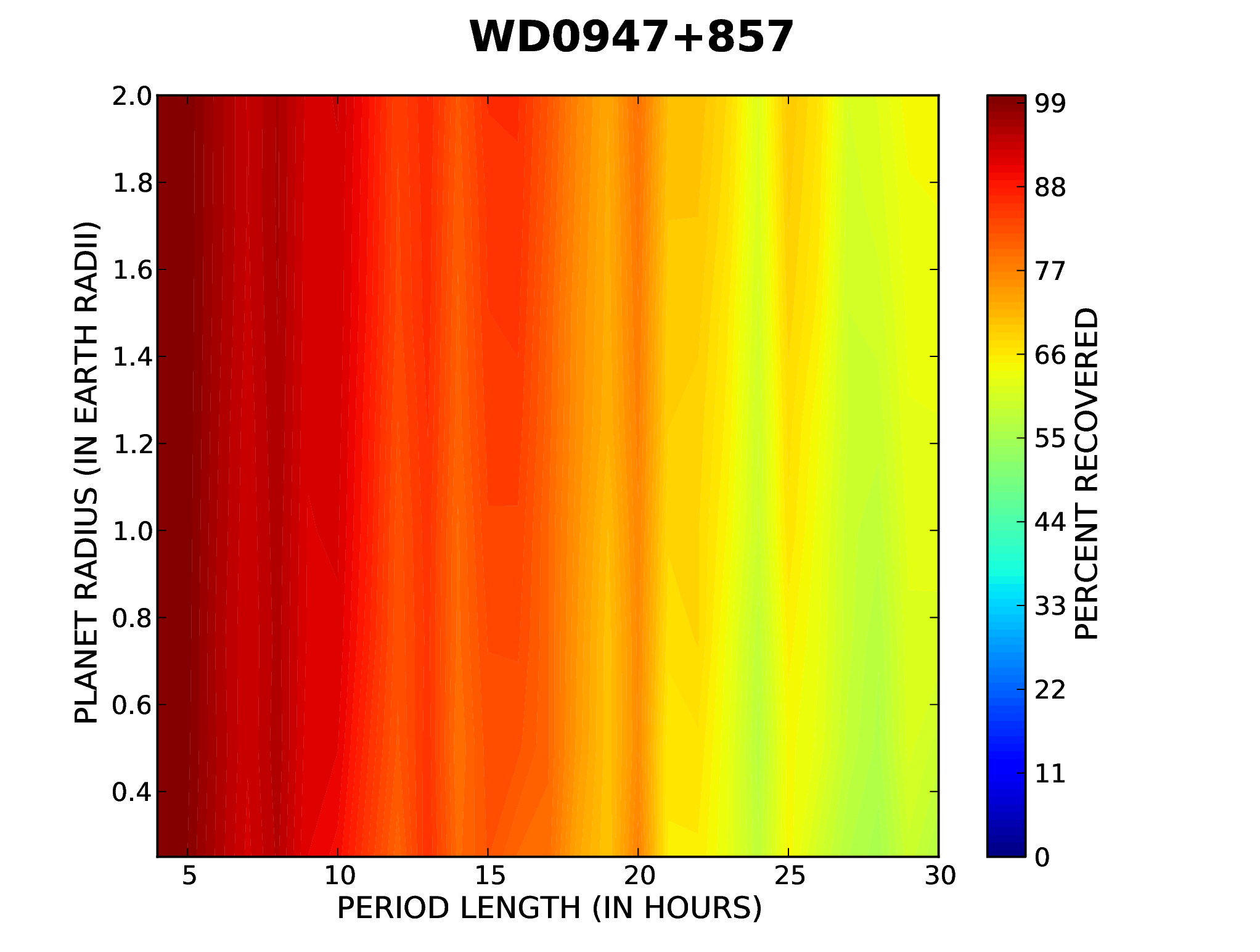}{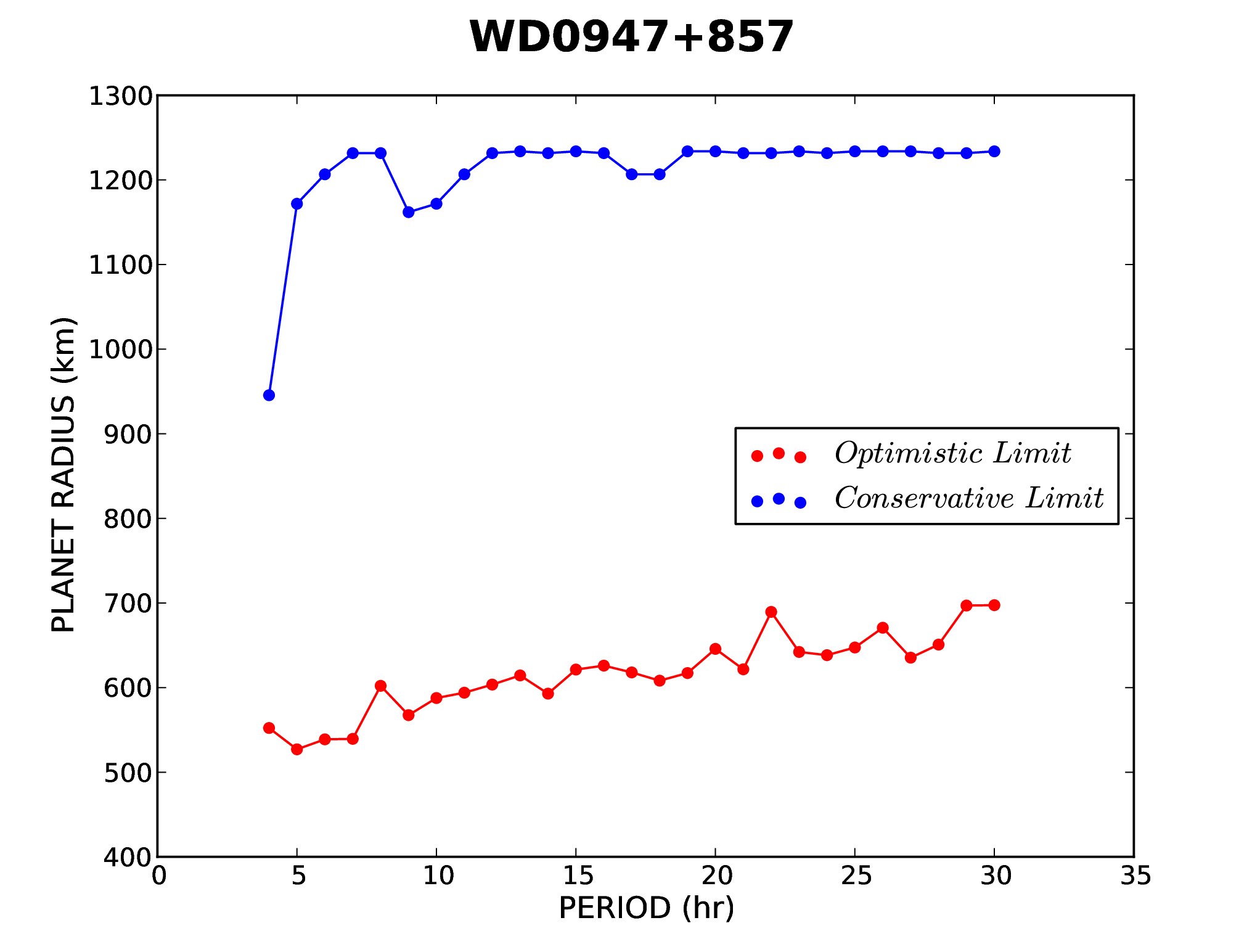}
\caption{\label{fig:0947} Same as Figure \ref{fig:0308} for WD 0947+857.}
\end{figure}

\begin{figure}
\plottwo{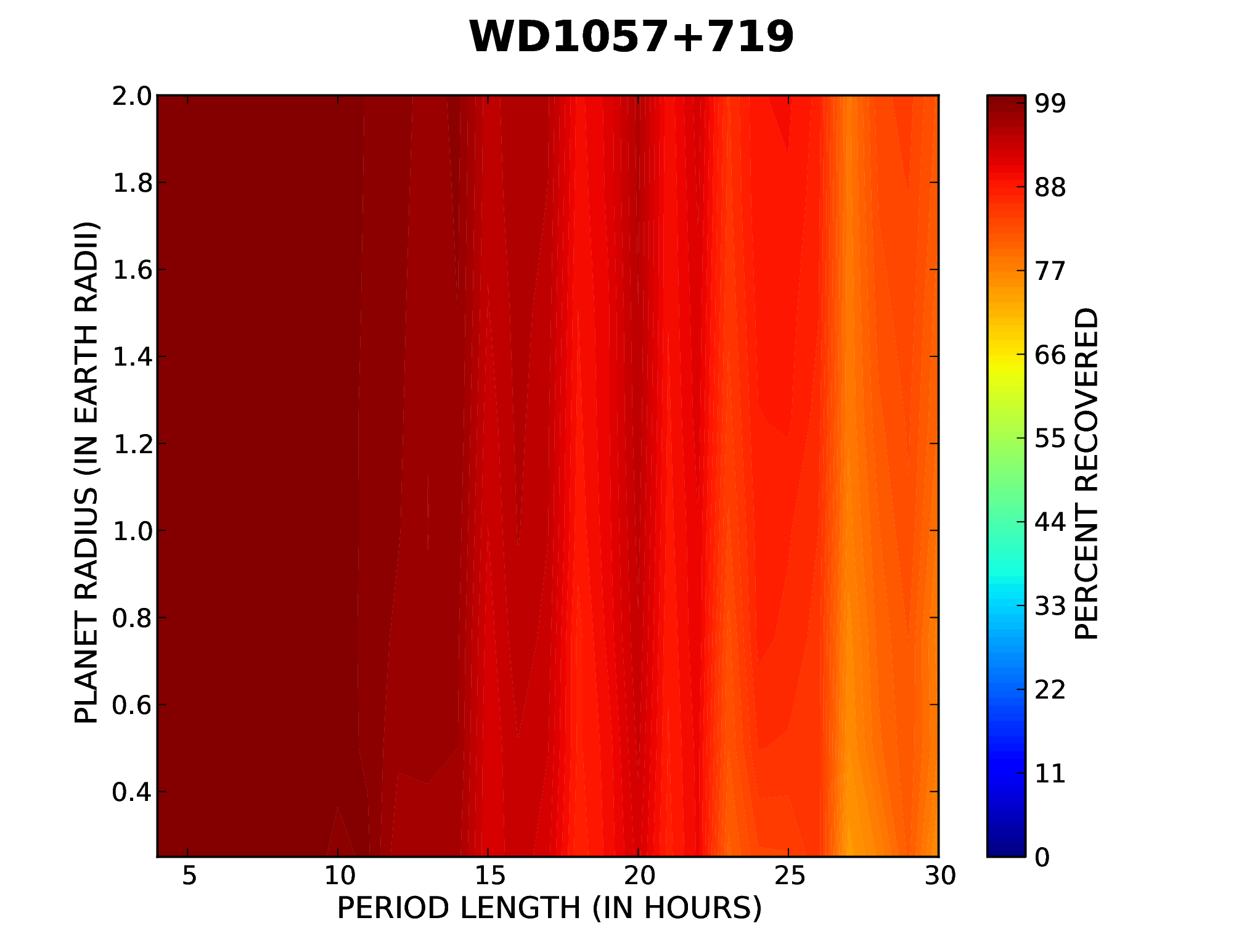}{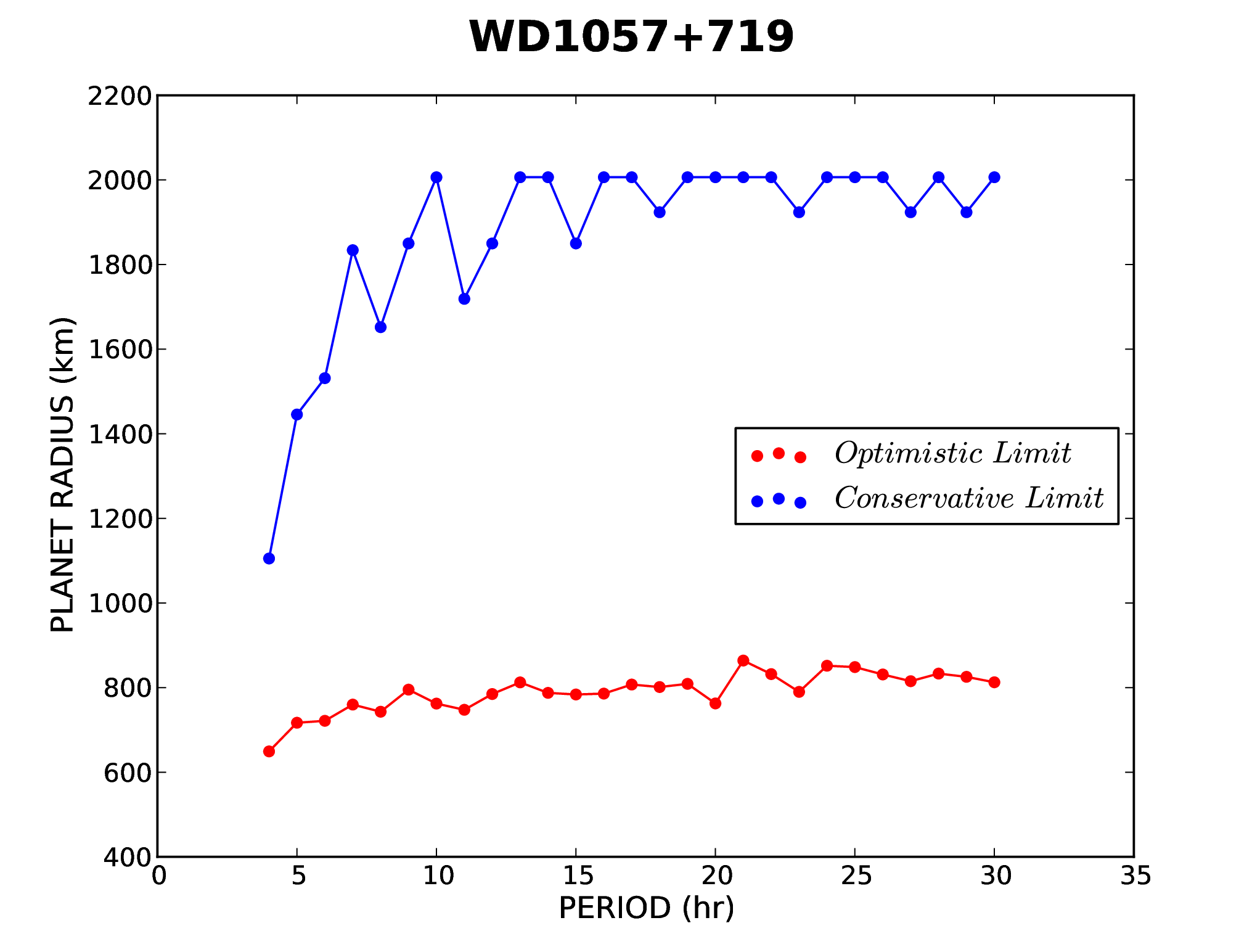}
\caption{\label{fig:1057} Same as Figure \ref{fig:0308} for WD 1057+719.}
\end{figure}

\begin{figure}
\plottwo{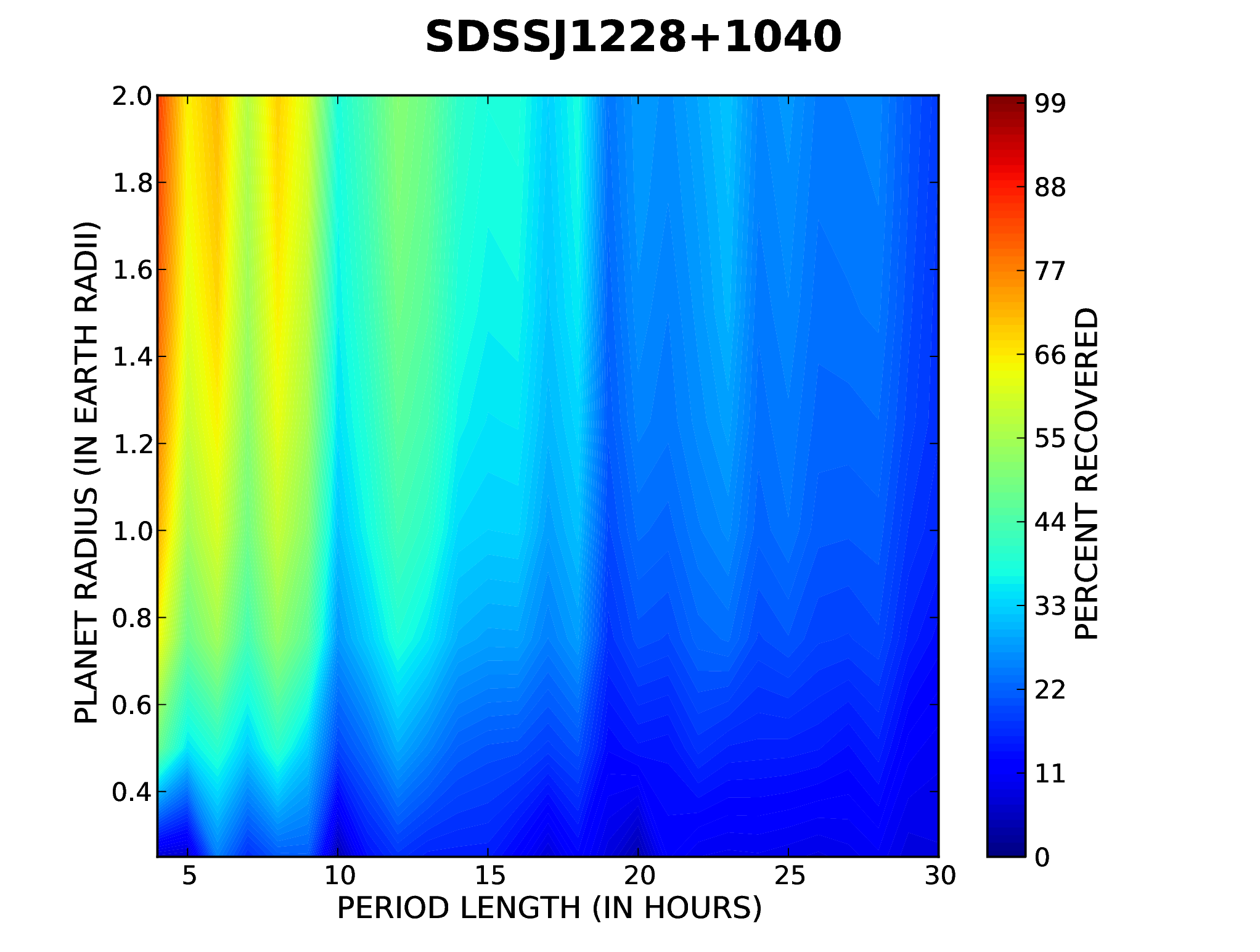}{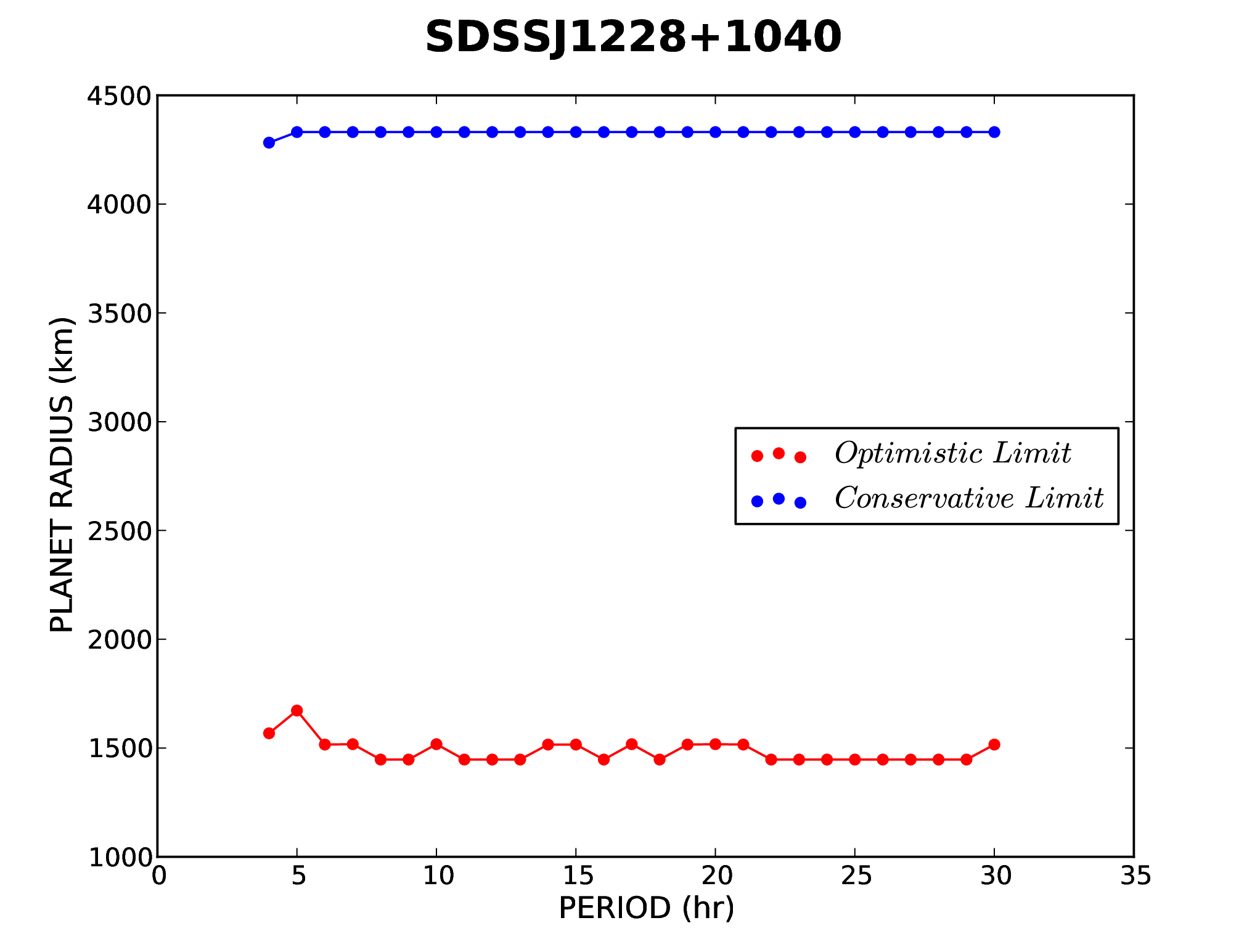}
\caption{\label{fig:1228} Same as Figure \ref{fig:0308} for SDSSJ 1228+1040.}
\end{figure}
\clearpage

\begin{figure}
\plotone{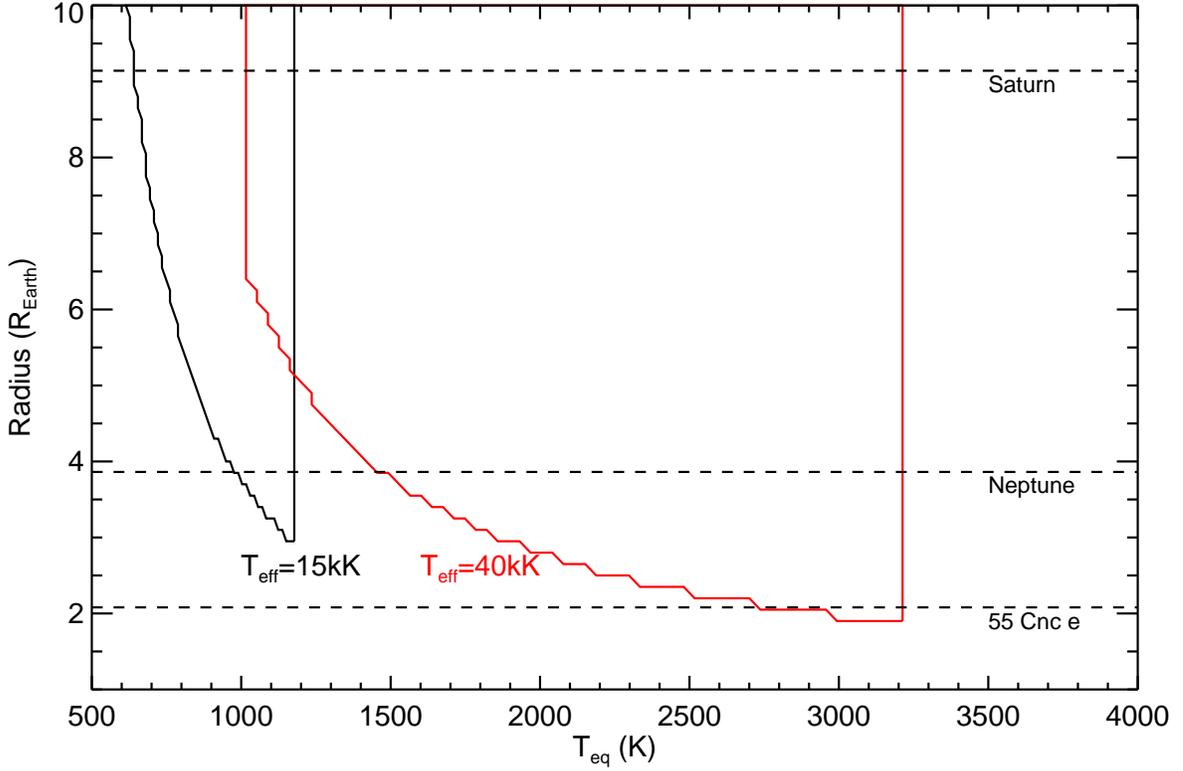}
\caption{\label{fig:superearth} We present theoretical radius limit curves of detectable irradiated sub-Jovian radius planets around two WDs with T$_{\mathrm eff}$=15000~K and 40000~K.  We calculate the limiting radii for companion orbits that span from 1~\Rsun\ to 10~\Rsun.  We assume the planets are in radiative thermal equilibrium with the stellar insolation.  In reality, the excesses may be more complex due to the atmosphere of the companion and day/night variations.  To determine the IR sensitivity we assume a $>$10\% excess in the W1 or W2 channel of WISE.}
\end{figure}

\begin{figure}
\plotone{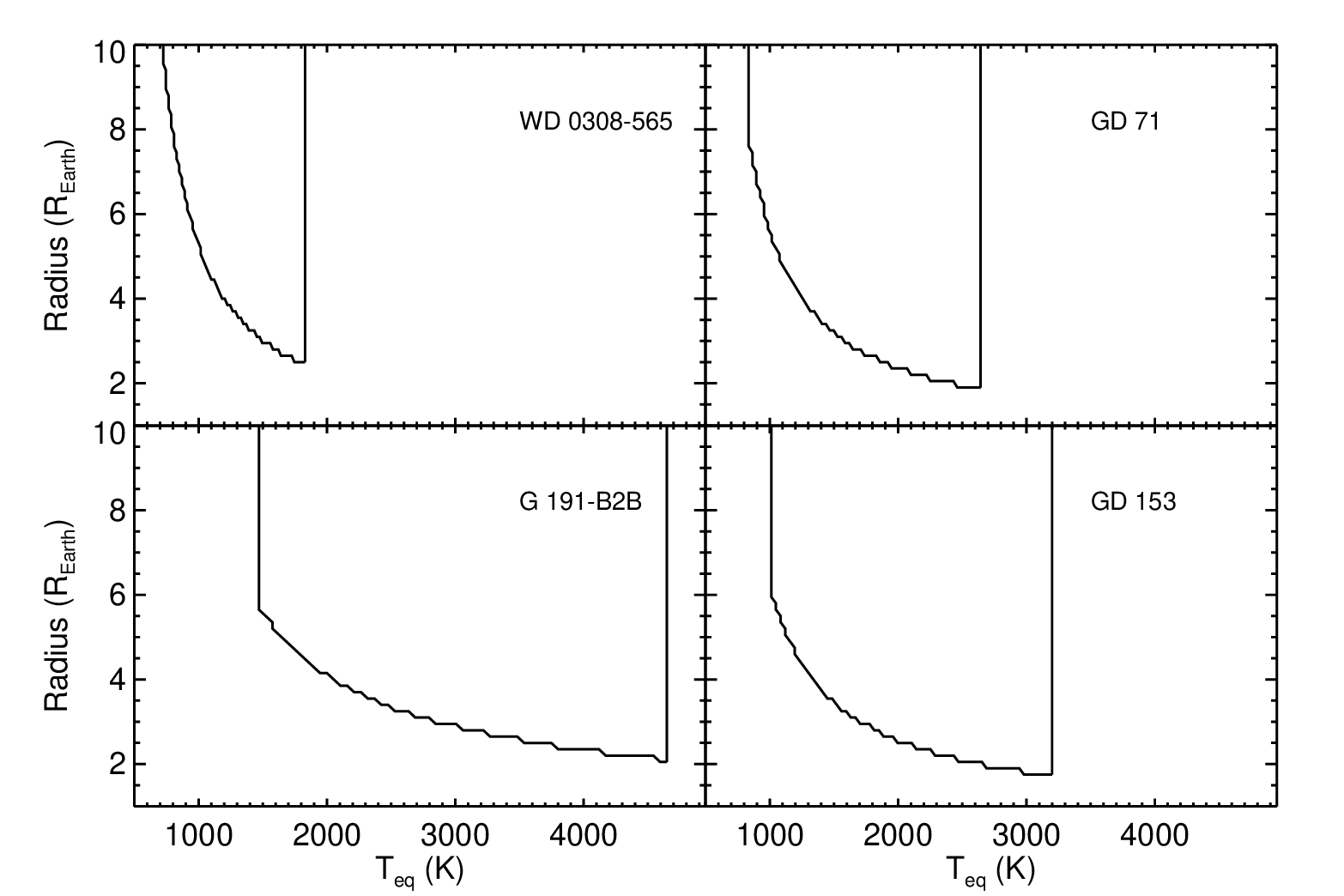}
\caption{\label{fig:irr1}  The same as Figure \ref{fig:superearth} but for four of our COS target WDs.}
\end{figure}

\begin{figure}
\plotone{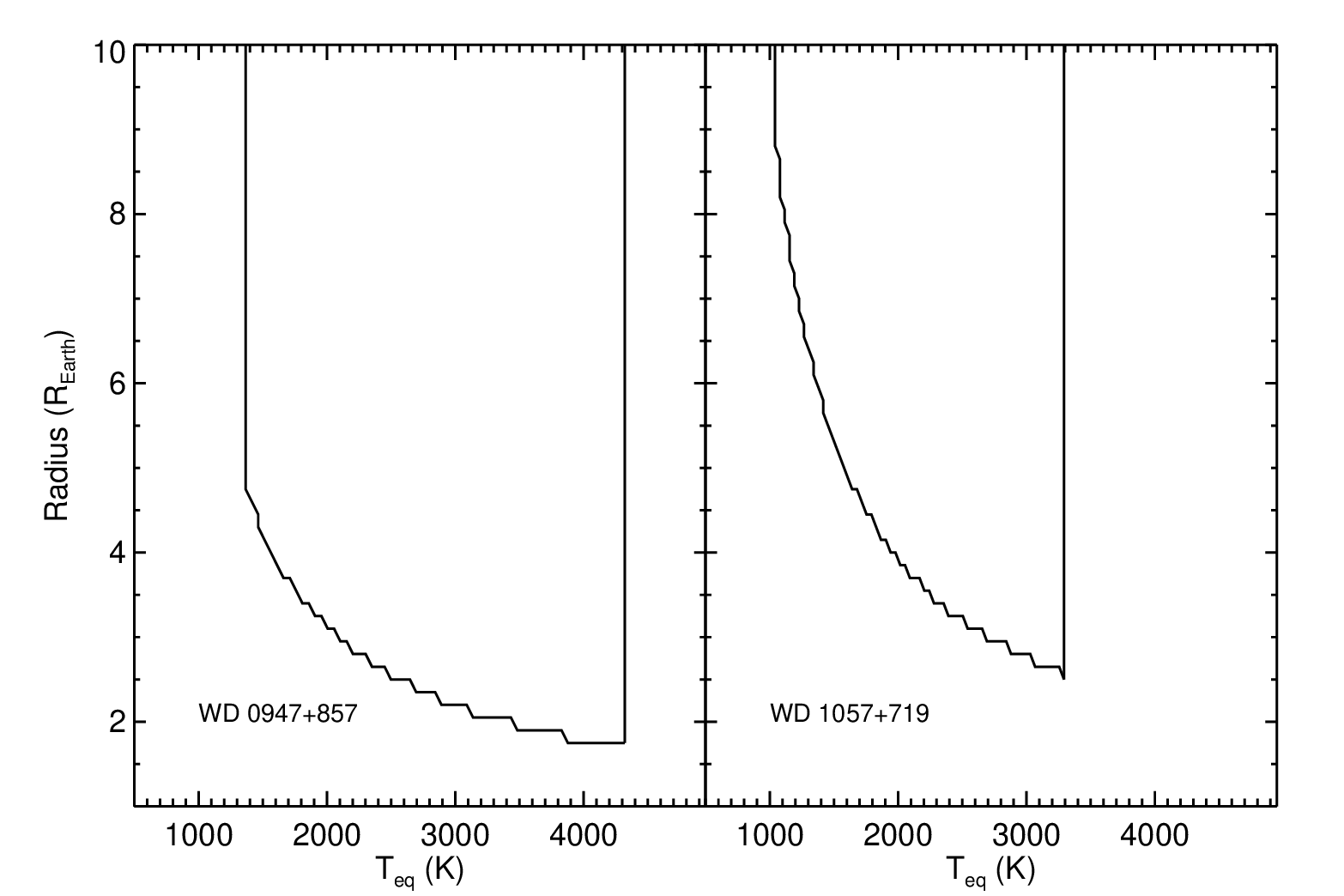}
\caption{\label{fig:irr2}The same as Figure \ref{fig:superearth} but for two of our COS target WDs.}
\end{figure}

\begin{figure}
\plotone{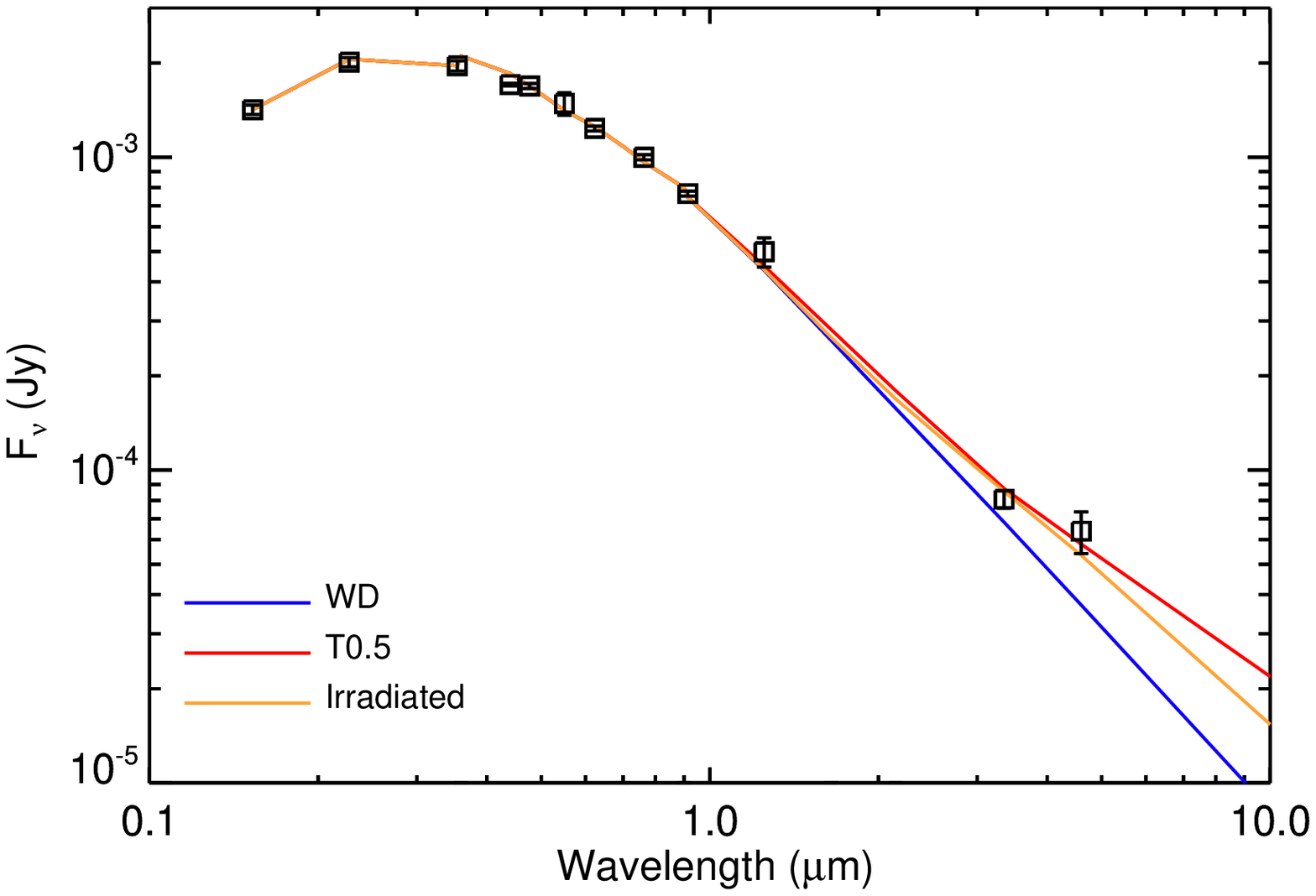}
\caption{\label{fig:024} SED of WD 0249-052 compared with possible companions responsible for the IR excess reported in H13.  The black squares are the observed photometry, while the blue solid line shows the expected WD photosphere.  The W1 and W2 points are both marginally in excess of the expected photosphere, and can be explained either by a faint brown dwarf with a T0-T1 spectral type, or by a close-in irradiated companion with a radius of 5.2~R$_\oplus$ and a T$_{\rm eff}$=1400~K.}
\end{figure}

\begin{figure}
\plotone{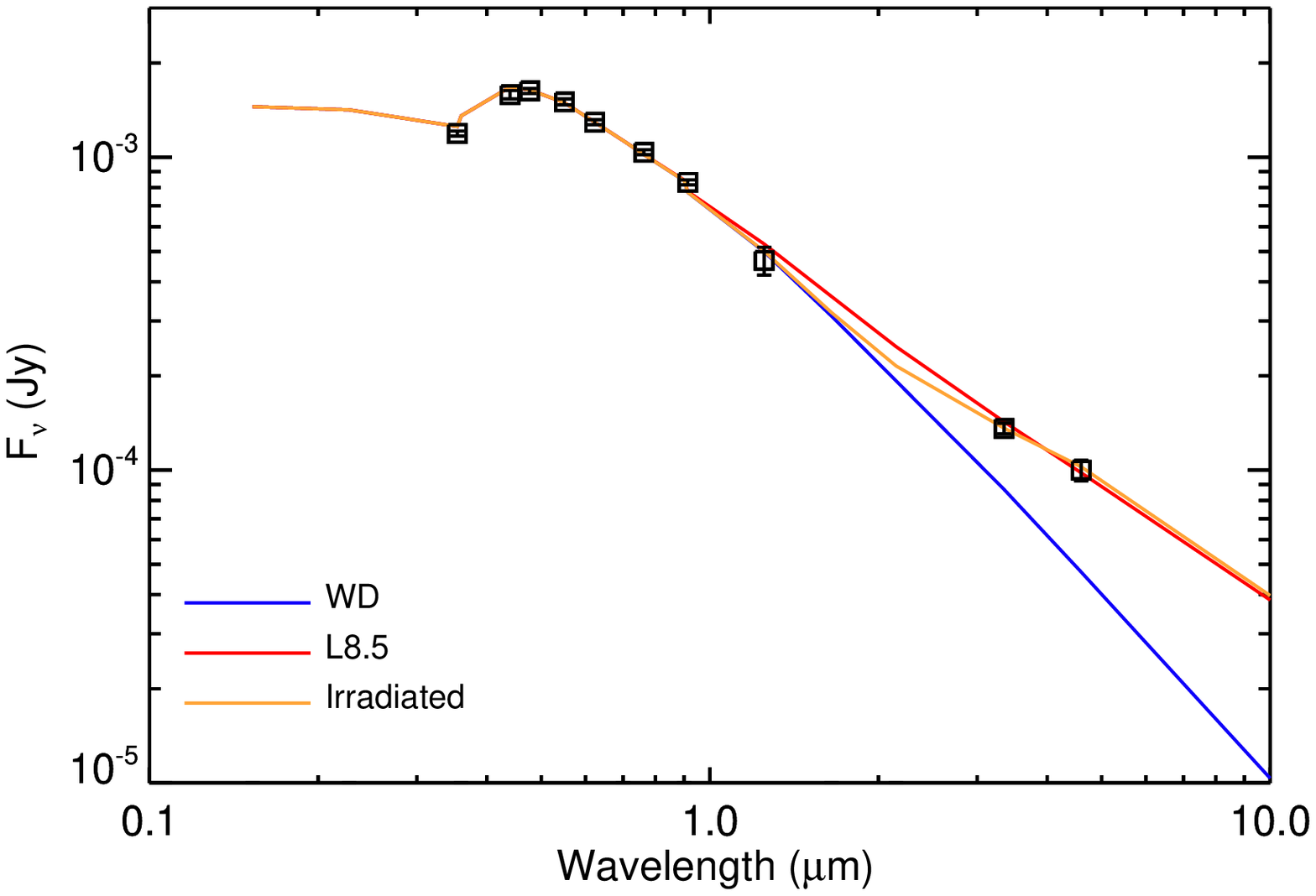}
\caption{\label{fig:144}SED of WD 1448+411 compared with possible companions responsible for the IR excess reported in H13.  The black squares are the observed photometry, while the blue solid line shows the expected WD photosphere.  The W1 and W2 points are significantly in excess of the expected photosphere, and can be explained either by a faint brown dwarf with a L8.5 spectral type, or by a close-in irradiated companion with a radius of 9~\Rearth and a T$_{\rm eff}$=1140~K.}
\end{figure}

\end{document}